%% file: main.tex
\documentclass[12pt,a4paper]{article}
\usepackage[left=2.5cm,right=2.5cm,top=3.5cm,bottom=3.5cm,a4paper]{geometry}
\pdfoutput=1 

\usepackage[T1]{fontenc} 
\usepackage[utf8]{inputenc}

\usepackage{amsmath,amssymb,xspace,hyperref,graphicx,color,ifpdf,ifthen,mciteplus}
\usepackage{booktabs,multirow,cite,braket}
\usepackage{bm} 
\usepackage{tikz}
\graphicspath{{sensitivity_figs/}} 

\title{\boldmath \textbf{Extracting maximum information from polarised baryon decays via amplitude analysis: the \Lcpkpi case}}

 \author{\textbf{Daniele Marangotto}\footnote{E-mail: \texttt{daniele.marangotto@unimi.it}}\\[2ex]Universit\`a degli Studi di Milano and INFN Milano,\\[2ex]Via Celoria 16, 20133 Milano, Italy}

\newboolean{pdflatex}
\setboolean{pdflatex}{true} 

\newboolean{articletitles}
\setboolean{articletitles}{true} 

\newboolean{uprightparticles}
\setboolean{uprightparticles}{false} 

\newboolean{inbibliography}
\setboolean{inbibliography}{false} 

\include{thesis-symbols}
\include{preamble}

\begin{document} 
\maketitle

\section*{Abstract}
We consider what is the maximum information measurable from the decay distributions of polarised baryon decays via amplitude analysis in the helicity formalism. We focus in particular on the analytical study of the \Lcpkpi decay distributions, demonstrating that the full information on its decay amplitudes can be extracted from its distributions, allowing a simultaneous measurement of both helicity amplitudes and the polarisation vector.
This opens the possibility to use the \Lcpkpi decay for applications ranging from New Physics searches to low-energy QCD studies, in particular its use as absolute polarimeter for the \Lc baryon.
This result is valid as well for baryon decays having the same spin structure and it is cross-checked numerically by means of a toy amplitude fit with Monte Carlo pseudo-data.

\clearpage

\tableofcontents

\clearpage

\section{Introduction}
\label{sec:intro}
The study of the complete phase space distributions (\ie the fully differential decay rate) of particle decays via angular or amplitude analysis allows to extract the maximum information about the process, since no integration is performed on the decay degrees of freedom. However, what is this maximum information for a given decay structure? Which parameters describing the phase space distributions can be measured? Indeed, in general, it is not guaranteed that the different functional forms characterising the decay distributions, separable by means of an amplitude fit, yield enough constraints on the parameters describing the decay in a given phenomenological framework. In this article we study the constraints placed by the phase space distributions of baryon decays described in the helicity formalism, showing what information can be obtained under which conditions.

In particular, we focus on the \Lcpkpi decay, whose amplitude analysis is ongoing at the LHCb experiment~\cite{Marangotto:2713231}, demonstrating the possibility to extract the full information on its parameters, measuring both helicity amplitudes and the polarisation vector simultaneously, in presence of non-negligible polarisation. Thus, the \Lcpkpi decay parameters can be considered as physical observables. This result is valid as well for polarised baryon decays having the same spin structure, a first parity-violating decay and a subsequent parity-conserving decay. Since the present article is intended to be a phenomenological study, we will not consider experimental effects: we assume a sufficiently large fit statistics, allowing to effectively separate each phase space dependency, and an adequate description of the invariant mass lineshape functions which parametrise resonant contributions.

The possibility to extract the whole decay amplitude is a remarkable result given the strong interest in the measurement of the associated observables, ranging from New Physics searches to low-energy QCD studies. The full knowledge of the helicity amplitudes characterises each resonant contribution to the decay, both its squared modulus and phase, as well the resonance polarisation. The comparison between observables measured for $C\!P$ conjugated decays enables $C\!P$ symmetry violation studies for specific contributions or localised in the phase space.
Moreover, the knowledge of a particle decay amplitude allows to add information on its production processes; for instance, the inclusion of the \Lcpkpi amplitude model in $\Lb\to\Lc l^-\bar{\nu}_l$ angular analyses increases the sensitivity to possible beyond the Standard Model physics contributions~\cite{Dutta:2015ueb,*Shivashankara:2015cta,*Li:2016pdv,*Datta:2017aue,*DiSalvo:2018ngq,*Ray:2018hrx,*Boer:2019zmp,*Penalva:2019rgt}.

Considering baryon polarisation, the measurement of its absolute value and direction is essential for a variety of studies. Polarisation measurements for different production mechanisms give precious information on the baryon spin structure and formation process; for heavy baryons they are expected to be closely related to the charm quark polarisation and its originating process~\cite{Mannel:1991bs,*Falk:1993rf,*Galanti:2015pqa}. Since the baryon polarisation is difficult to predict in QCD, being related to its non-perturbative regime, such measurements are useful to discriminate among different low energy QCD models.

Focusing on the \Lc baryon, its main decay channel \Lcpkpi allows the measurement of its polarisation with the best statistical precision.
Indeed, the two-body decay $\Lc\to \Lz\pi^+$, which can be used for polarisation measurements since its decay asymmetry parameter is known~\cite{PDG2018}, has a lower branching fraction by a factor $\approx 5$~\cite{PDG2018} and reduced detector reconstruction efficiency because of the large $\Lz$ baryon flight distance, especially for fixed-target experiments. Moreover, the use of single resonant components of the \Lcpkpi decay is not feasible because of its complicated decay structure characterised by many overlapping and interfering resonant contributions, making single components hardly to isolate~\cite{Marangotto:2713231}.

A method to extract the \Lc polarisation with the best precision is fundamental for the proposed search of charm baryon electromagnetic dipole moments using bent crystals at the  LHC~\cite{Botella:2016ksl, *Bagli:2017foe}, since dipole moments are to be inferred from spin precession.

A measurement of the \Lc polarisation in the \Lcpkpi decay mode via amplitude analysis has already been performed by the E791 experiment~\cite{Aitala:1999uq}; however, the results obtained are not reliable: first, because a wrong amplitude model was employed, since no matching of proton spin states among different decay chains was performed; second, because no analytical or numerical study showed where the sensitivity to the polarisation came from. This study addresses for the first time the question for \Lcpkpi decays.

The decay distribution of different non-leptonic \Lc decays has been studied theoretically~\cite{Korner:1992wi,Bialas:1992ny,Konig:1993wz}, also in connection with weak $\Lb\to\Lc$ transitions~\cite{Konig:1993ze}. A precise determination of the \Lcpkpi amplitude model would allow to test some theoretical predictions, for instance the parity-conserving nature of the $\Lambda_c \to \Delta^{++}K^-$ decay\cite{Korner:1992wi}. The \Lc longitudinal polarisation has been theoretically explored under $SU(3)$ flavor symmetry in Ref.~\cite{Cen:2019ims}.

We first introduce the formalism employed for the general expression of polarised decay rates in the helicity formalism, Section~\ref{sec:formalism}. We review the decay distributions of two-body and three-body via a single intermediate state baryon decays in Sections~\ref{sec:two-body_decays} and~\ref{sec:three-body_decays}, respectively, stressing the role of the baryon polarisation in the determination of the decay rate. In the latter case, we consider the decay distributions associated to the spin structure $1/2\to S_R(\to 1/2,0),0$, which is relevant \eg for $\Lc\to\Lz(\to p\pi^-) \pi^+$ decays.

The core of the article is the study of the decay rate of three-body decays via multiple intermediate states, for the \Lcpkpi case, Section~\ref{sec:Lc2pKpi}. We show how the presence of significant interference effects together with a non-negligible \Lc polarisation allows the simultaneous measurement of all the parameters characterising the \Lcpkpi amplitude model, including complex helicity couplings and the polarisation vector. In Section~\ref{sec:toy_fit} we cross-check the analytical study of the \Lcpkpi decay rate by means of a toy amplitude fit on Monte Carlo generated pseudo-data. The conclusions of the article are summarized in Section~\ref{sec:conclusions}.

\section{Formalism}
\label{sec:formalism}

\subsection{Polarised Differential Decay Rate}
\label{sec:polarised_rate}
We consider the differential decay rate for polarised particles, see \eg Ref.~\cite{Leader2011} for a more complete treatment of the subject. The generic spin state of a statistical ensemble of particles is described by means of a density operator: given an ensemble of spin states $\Ket{\psi}_i$ occurring with probability $p_i$, the density operator is
\begin{equation}
\hat{\rho} = \sum_i p_i \Ket{\psi}_i\Bra{\psi}_i,
\label{eq:density_matrix}
\end{equation}
and the expectation value of any operator $\hat{X}$ on the state described by $\hat{\rho}$ can be expressed as
\begin{equation}
\langle\hat{X}\rangle = \sum_i p_i \Braket{\psi|\hat{X}|\psi}_i = \mathrm{Tr}\left[\hat{\rho}\hat{X}\right].
\end{equation}

The decay rate of a multi-body decay $A\to \lbrace i=1,..,n \rbrace$ for definite spin eigenstates is the squared modulus of the transition amplitude between the $A$ particle initial state $\Ket{s_A,m_A}$ and the final particle product state $\Ket{\lbrace s_i\rbrace,\lbrace m_i\rbrace} = \otimes_i \Ket{s_i,m_i}$,
\begin{align}
p_{m_A,\lbrace m_i\rbrace}(\Omega) &= |\braket{s_A,m_A|\hat{T}|\lbrace s_i\rbrace,\lbrace m_i\rbrace}|^2 \nonumber\\
&= |\mathcal{A}_{m_A,\lbrace m_i\rbrace}(\Omega)|^2.
\label{eq:decay_rate_pure_states}
\end{align}
The label $\Omega$ denotes the set of phase space variables describing the decay distributions.

Generic polarisation states are described by introducing the density operators for the initial particle state $\hat{\rho}^A$ and the final particle state $\hat{\rho}^{\lbrace i\rbrace}$, which are included in the decay rate Eq.~\eqref{eq:decay_rate_pure_states} by inserting suitable identity resolutions,
\begin{align}
p(\Omega,\hat{\rho}^A,\hat{\rho}^{\lbrace i\rbrace}) &= \mathrm{tr} \left[ \hat{\rho}^A \hat{T} \hat{\rho}^{\lbrace i\rbrace} \hat{T}^{\dagger} \right]\nonumber\\
&= \sum_{m_A,m'_A} \sum_{\lbrace m_i\rbrace,\lbrace m'_i\rbrace} \hat{\rho}^A_{m_A,m'_A} \hat{\rho}^{\lbrace i\rbrace}_{\lbrace m_i\rbrace,\lbrace m'_i\rbrace} \mathcal{A}_{m_A,\lbrace m_i\rbrace}(\Omega) \mathcal{A}^*_{m'_A,\lbrace m'_i\rbrace}(\Omega).
\label{eq:decay_rate_mixed_states}
\end{align}

The differential decay rate of a spin 1/2 particle in a generic polarisation state is derived considering its density matrix
\begin{equation}
\rho^{A} = \frac{1}{2}\left(\mathbb{I} + \bm{P}\cdot\bm{\sigma}\right) = \frac{1}{2}
\left(
\begin{array}{cc}
1+P_z & P_x-iP_y\\
P_x+iP_y & 1-P_z\\
\end{array}
\right),
\label{eq:density_matrix_1/2}
\end{equation}
in which the polarisation components $P_x$, $P_y$, $P_z$ are the expectation values of the three spin operators and $\bm{\sigma}$ the three Pauli matrices. The final particle polarisation states, assumed to be unmeasurable, are described by an identity density matrix
\begin{equation}
\rho^{\lbrace i\rbrace} = \frac{\mathbb{I}}{2}.
\end{equation}

The differential rate Eq.~\eqref{eq:decay_rate_mixed_states}, decomposed into unpolarised, longitudinal ($P_z$) and orthogonal ($P_x$, $P_y$) polarisation parts becomes
\begin{align}
p(\Omega,\bm{P}) &= p_{unpol}(\Omega) + p_{long}(\Omega,P_z) + p_{orth}(\Omega,P_x,P_y),\\
p_{unpol}(\Omega) &= \frac{1}{2}\sum_{\lbrace m_i\rbrace} \left( |\mathcal{A}_{1/2,\lbrace m_i\rbrace}(\Omega)|^2 + |\mathcal{A}_{-1/2,\lbrace m_i\rbrace}(\Omega)|^2 \right),\label{eq:unpolarised_decay_rate}\\
p_{long}(\Omega,P_z) &= \frac{1}{2} P_z \sum_{\lbrace m_i\rbrace} \left( |\mathcal{A}_{1/2,\lbrace m_i\rbrace}(\Omega)|^2 - |\mathcal{A}_{-1/2,\lbrace m_i\rbrace}(\Omega)|^2 \right),\label{eq:longitudinal_pol_decay_rate}\\
p_{orth}(\Omega,P_x,P_y) &= \re \left[(P_x-iP_y) \sum_{\lbrace m_i\rbrace} \mathcal{A}_{1/2,\lbrace m_i\rbrace}(\Omega)\mathcal{A}^*_{-1/2,\lbrace m_i\rbrace}(\Omega) \right].\label{eq:orthogonal_pol_decay_rate}
\end{align}

\subsection{General Properties of the Polarised Decay Rate}
\label{sec:properties_polarised_decay_rate}
For later convenience we consider some properties of the polarised decay rate related to rotational invariance and parity symmetry.

The polarisation vector $\bm{P}$ associated to a decaying particle $A$ is the only quantity specifying a direction in its rest frame. Therefore, rotational invariance implies that for null polarisation the decay rate must be isotropic in any $A$ rest frame defined independently from the decay distributions. In other words, the decay rate specifies the relative angular distribution among daughter particles, but not their global orientation in space.

For non-zero polarisation both the polarisation vector and the daughter particle momenta transform in the same way under rotations. Thus, the relative orientation of daughter particles is independent on the polarisation vector.

The sensitivity of the decay rate to the particle polarisation depends critically on the amount of parity symmetry violation characterising the decay. Parity symmetry requires the decay angular distribution to be equal for $+|\bm{P}|$ and $-|\bm{P}|$ polarisation values, for any polarisation vector $\bm{P}$, since parity transformation reverses the daughter particle momenta but not the polarisation vector. Therefore, a decay mediated by a parity conserving interaction retains no information on the decaying particle polarisation.
Vice-versa, the sensitivity of the decay rate on parity-violating effects depends critically on the amount of decaying particle polarisation. Indeed, since for zero polarisation there is no preferred direction, the decay rate becomes symmetric under parity transformation, and parity-violating effects cancel.

One can see the combined effect of parity-violation and polarisation as creating an anisotropy along the direction specified by $\bm{P}$. Rotational invariance makes all such directions equivalent, in the sense that a rotation of the system can only change the direction of the anisotropy. Indeed, given a generic polarisation, we can choose the $z$ quantisation axis to be along $\bm{P}$ when studying the properties of the decay rate other than the polarisation direction. This is why in this article the sensitivity of the decay rate to its parameters like helicity couplings and polarisation modulus is usually studied assuming longitudinal polarisation only. Vice-versa, the anisotropy can be used to determine the polarisation direction from the decay distributions, as will be shown in Section~\ref{sec:two-body_decays}.

Note that, at the level of the observed decay distribution, parity-violation effects originated in the decay process may be influenced by final state interactions. A theoretical study would be needed to disentangle the two contributions from the results of an amplitude fit.

\subsection{Helicity Formalism}
\label{sec:helicity_formalism}
We briefly introduce the helicity formalism following the method of Ref.~\cite{Marangotto:2019ucc}, in which the helicity formalism is revisited in light of its application to multibody polarised particle decays, like the \Lcpkpi one. The ``standard'' helicity formalism of Ref.~\cite{JacobWick} is slightly modified to ease a correct matching of final particles spin states for decays featuring different interfering decay chains. For the case of two-body and three-body via a single intermediate state decays, Secs.~\ref{sec:two-body_decays} and~\ref{sec:three-body_decays}, where single decay amplitudes are involved, the two approaches coincide.

Two-body $A\to 1,2$ decay amplitudes can be expressed in terms of $A$ spin state $\ket{s_A,m_A}$, and a $1,2$ two-particle state being the product of particle 1 helicity states $\ket{s_1,\lambda_1}$ and particle 2 opposite-helicity states $\ket{s_2,\bar{\lambda}_2}$, as
\begin{align}
\mathcal{A}_{m_A,\lambda_1,\bar{\lambda}_2}(\theta_1,\phi_1) &= \Braket{\theta_1,\phi_1, \lambda_1,\bar{\lambda}_2|\hat{T}|s_A,m_A} \nonumber\\
&= \mathcal{H}_{\lambda_1,\bar{\lambda}_2} D^{*s_A}_{m_A,\lambda_1+\bar{\lambda}_2}(\phi_1,\theta_1,0),
\label{eq:two_body_amplitude_corrected}
\end{align}
in which $\theta_1,\phi_1$ are the spherical angles of the particle 1 momentum in the $A$ reference system and $D$ is a Wigner $D$ matrix representing rotations on spin states (see \eg Ref.~\cite{Richman} for their definition and properties). The use of opposite-helicity states eases the control of phases arising from the helicity rotations. The complex number
\begin{equation}
\mathcal{H}_{\lambda_1,\bar{\lambda}_2} \equiv \Braket{s_A,m_A,\lambda_1,\bar{\lambda}_2|\hat{T}|s_A,m_A},
\label{eq:helicity_couplings}
\end{equation}
called helicity coupling, encodes the decay dynamics and can not depend on $m_A$ for rotational invariance. The helicity values allowed by angular momentum conservation are
\begin{equation}
|\lambda_1|\leq s_1,\hspace{1cm} |\bar{\lambda}_2|\leq s_2,\hspace{1cm} |\lambda_1+\bar{\lambda}_2|\leq s_A.
\label{eq:allowed_helicity_couplings}
\end{equation}

Multi-body decay amplitudes are treated in the helicity formalism by breaking the decay into sequences of two-body decays introducing suitable intermediate states and summing over their helicity states allowed by Eq.~\eqref{eq:allowed_helicity_couplings}.

\section{Two-body Decay}
\label{sec:two-body_decays}
We consider a two-body decay $A\to 1,2$ of a spin 1/2 particle in the helicity formalism introduced in Section~\ref{sec:helicity_formalism}. Following Eqs.~\eqref{eq:unpolarised_decay_rate},~\eqref{eq:longitudinal_pol_decay_rate}, the longitudinal polarisation decay rate is, 
\begin{equation}
p(\theta_1,P_z) = \sum_{\lambda_1,\bar{\lambda}_2} \left| \mathcal{H}_{\lambda_1,\bar{\lambda}_2} \right|^2 \left( \frac{1+P_z}{2} \text{ } d^{1/2}_{1/2,\lambda_1+\bar{\lambda}_2}(\theta_1)^2 + \frac{1-P_z}{2} \text{ } d^{1/2}_{-1/2,\lambda_1+\bar{\lambda}_2}(\theta_1)^2 \right),
\end{equation}
with $\theta_1,\phi_1$ are the spherical angles of the particle 1 momentum in the $A$ reference system.
The rate can not depend on the azimuthal angle $\phi_1$ for invariance under rotations around the $z$ axis. Using the $d$-matrix property
\begin{equation}
\sum_m d^{S}_{m,m'}(\theta)^2 = 1,
\end{equation}
and fixing the overall helicity coupling normalisation to
\begin{equation}
\sum_{\lambda_1,\bar{\lambda}_2} \left| \mathcal{H}_{\lambda_1,\bar{\lambda}_2} \right|^2 \equiv 1,
\label{eq:helicity_coupling_normalisation}
\end{equation}
we find that for zero polarisation the decay rate is constant, isotropic as required by rotational invariance. For non-zero polarisation the decay rate takes the well-known form
\begin{align}
p(\theta_1,P_z) &= \frac{1}{2} + \frac{P_z}{2} \sum_{\lambda_1,\bar{\lambda}_2} \left| \mathcal{H}_{\lambda_1,\bar{\lambda}_2} \right|^2 \left( d^{1/2}_{1/2,\lambda_1+\bar{\lambda}_2}(\theta_1)^2 - d^{1/2}_{-1/2,\lambda_1+\bar{\lambda}_2}(\theta_1)^2 \right) \nonumber\\
&= \frac{1}{2} + \frac{P_z}{2} \cos\theta_1 \sum_{\lambda_1,\bar{\lambda}_2} \text{sign}(\lambda_1+\bar{\lambda}_2) \left| \mathcal{H}_{\lambda_1,\bar{\lambda}_2} \right|^2\nonumber\\
&= \frac{1}{2}\left(1 + \alpha P_z \cos\theta_1 \right),
\label{eq:two_body_decay_rate}
\end{align}
in which the explicit expression for the $d$-matrices Eq.\eqref{eq:d_matrix_1/2} has been used and the decay asymmetry parameter
\begin{equation}
\alpha \equiv \sum_{\lambda_1,\bar{\lambda}_2} \text{sign}(\lambda_1+\bar{\lambda}_2) \left| \mathcal{H}_{\lambda_1,\bar{\lambda}_2} \right|^2,
\end{equation}
is introduced.
The sensitivity to the polarisation is governed by the $\alpha$ parity-violating parameter. Indeed $\alpha$ is zero if the decay conserves parity, which requires
\begin{equation}
\left| \mathcal{H}_{\lambda_1,\bar{\lambda}_2} \right|^2 = \left| \mathcal{H}_{-\lambda_1,-\bar{\lambda}_2} \right|^2.
\end{equation}
Note that a fit to the $\cos\theta_1$ decay distribution can only measure the combination $\alpha P_z$: it is not possible to determine separately the polarisation and the $\alpha$ parameter values, unless one of the two is available from other measurements. Moreover, the fit is not sensitive to the single helicity couplings, but only to the $\alpha$ combination.

For a generic $A$ polarisation vector, the decay rate is, following Eqs.~\eqref{eq:unpolarised_decay_rate}~\eqref{eq:longitudinal_pol_decay_rate},~\eqref{eq:orthogonal_pol_decay_rate},
\begin{align}
p(\theta_1,\phi_1,\bm{P}) = \frac{1}{2} \sum_{\lambda_1,\bar{\lambda}_2} \left| \mathcal{H}_{\lambda_1,\bar{\lambda}_2} \right|^2 &\left[ (1+P_z) \text{ } d^{1/2}_{1/2,\lambda_1+\bar{\lambda}_2}(\theta_1)^2 + (1-P_z) \text{ } d^{1/2}_{-1/2,\lambda_1+\bar{\lambda}_2}(\theta_1)^2 \right.  \nonumber\\
&+\left. 2 \left( P_x\cos\phi_1 + P_y\sin\phi_1 \right) d^{1/2}_{1/2,\lambda_1+\bar{\lambda}_2}(\theta_1) \text{ } d^{1/2}_{-1/2,\lambda_1+\bar{\lambda}_2}(\theta_1) \right].
\end{align}
The orthogonal polarisation part becomes
\begin{align}
p_{\rm orth}(\theta_1,\phi_1,P_x,P_y) &= \sum_{\lambda_1,\bar{\lambda}_2} \text{sign}(\lambda_1+\bar{\lambda}_2) \left| \mathcal{H}_{\lambda_1,\bar{\lambda}_2} \right|^2 \left( P_x\cos\phi_1 + P_y\sin\phi_1 \right) \cos\frac{\theta_1}{2}\sin\frac{\theta_1}{2} \nonumber\\
&=\frac{1}{2} \alpha \left( P_x\cos\phi_1 + P_y\sin\phi_1 \right) \sin\theta_1,
\end{align}
so that the decay rate is
\begin{align}
p(\theta_1,\phi_1,\bm{P}) &= \frac{1}{2}\left(1 + \alpha P_z \cos\theta_1 + \alpha P_x \sin\theta_1\cos\phi_1 + \alpha P_y \sin\theta_1\sin\phi_1\right) \nonumber\\
&= \frac{1}{2}\left(1 + \alpha \bm{P} \cdot \bm{\hat{p}}_1 \right),
\label{eq:two_body_decay_rate_generic_pol}
\end{align}
with $\bm{\hat{p}}_1$ being the particle 1 momentum versor in the $A$ reference system.
It features three angular distributions, each describing a different polarisation component, all multiplied by the $\alpha$ parameter. Therefore, a fit can determine the polarisation direction, but not its modulus independently of $\alpha$. Note that the orthogonal polarisation part does not add information on the helicity couplings, which enter the decay rate only via the $\alpha$ combination.

\section{Three-body Decay Via a Single Intermediate State}
\label{sec:three-body_decays}
We consider a three-body decay with a single intermediate state $R$ of the form $A\to R(\to 1,2),3$, with spin structure $1/2\to S_R(\to 1/2,0),0$. Summing over the resonance helicity states allowed by angular momentum conservation, the decay amplitude is,
\begin{equation}
\mathcal{A}_{m_A,\lambda_1} = \sum_{\lambda_R = \pm 1/2} \mathcal{H}^A_{\lambda_R,0} D^{*1/2}_{m_A,\lambda_R}(\phi_R,\theta_R,0) \mathcal{H}^R_{\lambda_1,0} D^{*S_R}_{\lambda_R,\lambda_1}(\phi_1,\theta_1,0) \mathcal{R}_{R}(m^2_R),
\end{equation}
in which $\theta_R,\phi_R$ are the spherical angles of the $R$ momentum in the $A$ reference system, $\theta_1,\phi_1$ now are the spherical angles of the particle 1 momentum in the $R$ helicity reference system, $\lambda_R$ is the $R$ helicity defined from the $A$ system and $\lambda_1$ is the particle 1 helicity defined from the $R$ helicity system. The helicity couplings $\mathcal{H}^A_{\lambda_R,0}$ and $\mathcal{H}^R_{\lambda_1,0}$ are associated to the two-body decays $A\to R,3$ and $R\to 1,2$, respectively. A non-negligible width for the $R$ state has been assumed, its invariant mass dependence described by the lineshape function $\mathcal{R}_{R}(m^2_R)$.
The squared modulus is
\begin{align}
\left|\mathcal{A}_{m_A,\lambda_1}\right|^2 &= \left[\left|\mathcal{H}^A_{1/2,0} \right|^2 \left|\mathcal{H}^R_{\lambda_1,0} \right|^2 d^{1/2}_{m_A,1/2}(\theta_R)^2 d^{S_R}_{1/2,\lambda_1}(\theta_1)^2\right. \nonumber\\
&+ \left|\mathcal{H}^A_{-1/2,0} \right|^2 \left|\mathcal{H}^R_{\lambda_1,0} \right|^2 d^{1/2}_{m_A,-1/2}(\theta_R)^2 d^{S_R}_{-1/2,\lambda_1}(\theta_1)^2 \nonumber\\
&+ \left.2\re\hspace{-4pt}\left( \mathcal{H}^A_{1/2,0} \mathcal{H}^{*A}_{-1/2,0} e^{i\phi_1} \right)
d^{1/2}_{m_A,1/2}(\theta_R) d^{1/2}_{m_A,-1/2}(\theta_R) \left|\mathcal{H}^R_{\lambda_1,0}\right|^2  d^{s_R}_{1/2,\lambda_1}(\theta_1) d^{s_R}_{-1/2,\lambda_1}(\theta_1)\right]\nonumber\\
&\times |\mathcal{R}_{R}(m^2_R)|^2.
\end{align}
Let's consider the unpolarised decay rate Eq.~\eqref{eq:unpolarised_decay_rate}: using
\begin{equation}
\sum_{m_A = \pm 1/2} d^{1/2}_{m_A,\lambda}(\theta_R)^2 = 1, \hspace{1cm} \sum_{m_A = \pm 1/2} d^{1/2}_{m_A,1/2}(\theta_R) \text{ } d^{1/2}_{m_A,-1/2}(\theta_R) = 0,
\end{equation}
and the explicit $d$-matrix element values Eqs.~\eqref{eq:d_matrix_1/2},~\eqref{eq:d_matrix_s_1/2}, we find
\begin{align}
p_{\rm unpol}(m^2_R,\theta_1) &= \frac{1}{2} \left(\left|\mathcal{H}^A_{1/2,0} \right|^2 \left|\mathcal{H}^R_{1/2,0} \right|^2 + \left|\mathcal{H}^A_{-1/2,0} \right|^2 \left|\mathcal{H}^R_{-1/2,0} \right|^2\right) |g^{S_{R}}_{R}(m^2_R,\theta_1)|^2\cos^2\frac{\theta_1}{2} \nonumber\\
&+ \frac{1}{2} \left( \left|\mathcal{H}^A_{1/2,0} \right|^2 \left|\mathcal{H}^R_{-1/2,0} \right|^2 + \left|\mathcal{H}^A_{-1/2,0} \right|^2 \left|\mathcal{H}^R_{1/2,0} \right|^2 \right) |g^{S_{R}}_{R}(m^2_R,\theta_1)|^2\sin^2\frac{\theta_1}{2},
\end{align}
in which
\begin{equation}
g^{S_{R}}_{R}(m^2_R,\theta_1) \equiv f(S_R,\theta_1) \mathcal{R}_{R}(m^2_R),
\end{equation}
with $f(S_R,\theta_1)$ describing the angular distribution of the state $R$ for $S_R>1/2$, see Appendix~\ref{sec:d_matrix_expressions}. The rate can be arranged in the form (playing with the $\cos^2\alpha+\sin^2\alpha = 1$ relation)
\begin{equation}
p_{\rm unpol}(m^2_R,\theta_1) = \frac{1}{4} \left( 1+ \alpha_A \alpha_R \cos\theta_1 \right) |g^{S_{R}}_{R}(m^2_R,\theta_1)|^2,
\end{equation}
by using the normalisation condition Eq.~\eqref{eq:helicity_coupling_normalisation} and the definition of the asymmetry parameters for the two-body decays $A\to R,3$ and $R\to 1,2$,
\begin{equation}
\alpha_A \equiv \left|\mathcal{H}^A_{1/2,0} \right|^2 - \left|\mathcal{H}^A_{-1/2,0} \right|^2, \hspace{1cm}
\alpha_R \equiv \left|\mathcal{H}^R_{1/2,0} \right|^2 - \left|\mathcal{H}^R_{-1/2,0} \right|^2.
\end{equation}
A fit to the $\cos\theta_1$ distribution determines the combination $\alpha_A \alpha_R$, and the two can not be separated unless one of the two is already measured. Note that the different form of the $\cos\theta_1$ distribution for different $S_R$ values can be exploited to measure the $R$ spin if not known.

The longitudinal polarisation decay rate Eqs.~\eqref{eq:unpolarised_decay_rate},~\eqref{eq:longitudinal_pol_decay_rate}, applying
\begin{equation}
d^{1/2}_{1/2,\lambda}(\theta_R)^2 - d^{1/2}_{-1/2,\lambda}(\theta_R)^2 = \cos^2\frac{\theta_R}{2} - \sin^2\frac{\theta_R}{2} = \cos\theta_R,
\end{equation}
and
\begin{equation}
d^{1/2}_{1/2,1/2}(\theta_R) d^{1/2}_{1/2,-1/2}(\theta_R) - d^{1/2}_{-1/2,1/2}(\theta_R) d^{1/2}_{-1/2,-1/2}(\theta_R) = -2\cos\frac{\theta_R}{2}\sin\frac{\theta_R}{2} = -\sin\theta_R
\end{equation}
becomes equal to
\begin{align}
p(m^2_R,\theta_R,&\theta_1,\phi_1,P_z) = p_{\rm unpol}(\theta_1)\nonumber\\
&+ \frac{1}{2} P_z \cos\theta_R \left(\left|\mathcal{H}^A_{1/2} \right|^2 \left|\mathcal{H}^R_{1/2} \right|^2 - \left|\mathcal{H}^A_{-1/2} \right|^2 \left|\mathcal{H}^R_{-1/2} \right|^2\right) |g^{S_{R}}_{R}(m^2_R,\theta_1)|^2 \cos^2\frac{\theta_1}{2} \nonumber\\
&+ \frac{1}{2} P_z \cos\theta_R \left( \left|\mathcal{H}^A_{1/2} \right|^2 \left|\mathcal{H}^R_{-1/2} \right|^2 - \left|\mathcal{H}^A_{-1/2} \right|^2 \left|\mathcal{H}^R_{1/2} \right|^2 \right) |g^{S_{R}}_{R}(m^2_R,\theta_1)|^2\sin^2\frac{\theta_1}{2}\nonumber\\
&+ \frac{1}{2} P_z \re\left( \mathcal{H}^A_{1/2} \mathcal{H}^{*A}_{-1/2} e^{i\phi_1} \right) \left|\mathcal{H}^R_{1/2}\right|^2 (-\sin\theta_R) |g^{S_{R}}_{R}(m^2_R,\theta_1)|^2 \sin\theta_1
\nonumber\\
&+ \frac{1}{2} P_z \re\left( \mathcal{H}^A_{1/2} \mathcal{H}^{*A}_{-1/2} e^{i\phi_1} \right) \left|\mathcal{H}^R_{-1/2}\right|^2 (\sin\theta_R) |g^{S_{R}}_{R}(m^2_R,\theta_1)|^2 \sin\theta_1.
\end{align}
In a similar way as before, the longitudinal polarisation decay rate can be written as
\begin{align}
p(m^2_R,\theta_R,\theta_1,\phi_1,P_z) &= \frac{1}{4} \left( 1+ \alpha_A \alpha_R \cos\theta_1 \right) |g^{S_{R}}_{R}(m^2_R,\theta_1)|^2 \nonumber\\
&+ \frac{1}{4} P_z\cos\theta_R \left( \alpha_A + \alpha_R \cos\theta_1 \right) |g^{S_{R}}_{R}(m^2_R,\theta_1)|^2 \nonumber\\
&- \frac{1}{2} P_z \alpha_R \left|\mathcal{H}^A_{1/2} \mathcal{H}^{*A}_{-1/2}\right| \cos(\phi_1+\Phi_A) \sin\theta_R |g^{S_{R}}_{R}(m^2_R,\theta_1)|^2 \sin\theta_1.
\label{eq:decay_rate_polarised_3body_decay}
\end{align}
in which $\Phi_A \equiv \mathrm{arg} \text{ }\mathcal{H}^A_{1/2} \mathcal{H}^{*A}_{-1/2}$ is the relative phase between the two $A\to R,3$ decay helicity couplings.

A fit to this decay distribution yields five constraints in total. The first two lines of Eq.~\eqref{eq:decay_rate_polarised_3body_decay} determine the three products $\alpha_A \alpha_R$, $P_z\alpha_A$, $P_z\alpha_R$: it is possible to separately extract $\alpha_A$, $\alpha_R$ and $P_z$ if each one is different from zero. If the intermediate state decay conserves parity, $\alpha_R = 0$, the decay distribution becomes analogous to that of the two-body $A\to R,3$ decay, Eq.~\eqref{eq:two_body_decay_rate},
\begin{equation}
p(m^2_R,\theta_R,\theta_1,P_z) = \frac{1}{2}\left(1 + P_z \alpha_A \cos\theta_R \right) |g^{S_{R}}_{R}(m^2_R,\theta_1)|^2,
\label{eq:decay_rate_3body_weak_strong_decay}
\end{equation}
for which $P_z$ and $\alpha_A$ can not be separately measured.

The last term of Eq.~\eqref{eq:decay_rate_polarised_3body_decay} allows the determination of $|\mathcal{H}^A_{1/2} \mathcal{H}^{*A}_{-1/2}|$ and $\Phi_A$ from the amplitude and phase of the $\phi_1$ oscillation, provided both $P_z$ and $\alpha_R$ are non-zero. Therefore, in presence of a sizeable polarisation and of two subsequent parity-violating decays, the helicity couplings associated to the $A\to R,3$ process are entirely measurable from the decay distributions, separately from the polarisation degree, since the decay rate gives 5 constraints on 5 real parameters (two for each complex coupling). In the following section, we will show how the same is valid for the \Lcpkpi decay as well, but for one difference: here, one of the crucial conditions is parity-violation in the $R$ decay; there, it is the interference among resonant contributions.

\section{Three-body Decay Via Multiple Intermediate States: the \Lcpkpi Case}
\label{sec:Lc2pKpi}
In this section, we study baryon three-body decays via multiple interfering intermediate states, considering parity-violating baryon decays to intermediate states decaying via parity-conserving interactions. We take \Lcpkpi decays as an explicit example but the results obtained in this section hold for any three-body decay with spin structure $1/2 \to 1/2 \text{ } 0 \text{ } 0$. The study of the \Lcpkpi decay distributions is complicated because of the different interfering decay chains, and it is split into the following parts.

First, we introduce the description of the three-body phase space and the Dalitz plot decomposition~\cite{Mikhasenko:2019rjf}, Section~\ref{sec:DPD}, which allows to separate invariant mass and decay orientation degrees of freedom in the decay amplitude. In Section~\ref{sec:general_study_pol} we perform a general study of the \Lcpkpi polarised decay rate, while in Section~\ref{sec:decay_amplitudes} we write the complete \Lcpkpi decay amplitude written in the helicity formalism following the method of Ref.~\cite{Marangotto:2019ucc}. The study of the \Lcpkpi rate is developed in Section~\ref{sec:decay_rate_study}.

\subsection{Three-body Decay Phase Space and Dalitz Plot Decomposition}
\label{sec:DPD}
A particle three-body decay is described by 5 degrees of freedom, resulting from 12 four momentum components constrained by 3 mass requirements and 4 energy-momentum conservation relations, which confine the daughters momenta to a plane in the rest frame of the mother particle. For unpolarised particles the decay plane orientation is irrelevant and the decay can be described by two two-body invariant masses (Dalitz variables). For the \Lcpkpi decay, \mqpk and \mqkpi are selected. For non-zero polarisation the orientation of the decay plane must be specified with respect to the \Lc spin coordinate system. The orientation can be expressed by means of the three Euler angles (introduced \eg in Ref.~\cite{Richman}) describing the rotation from the \Lc spin coordinate system to a decay plane reference system, chosen in such a way that the proton momentum defines the $z$ axis, while the component of the kaon momentum orthogonal to the proton momentum defines the $x$ axis,
\begin{align}
\bm{\hat{z}}_{\rm DP} = \bm{\hat{p}}(p), \hspace{1cm}
\bm{\hat{x}}_{\rm DP} = \frac{\bm{p}(p) \times \bm{p}(K^-)}{\left|\bm{p}(p) \times \bm{p}(K^-)\right|} \times \bm{\hat{p}}(p), \hspace{1cm}
\bm{\hat{y}}_{\rm DP} = \bm{\hat{z}}_{\rm DP} \times \bm{\hat{x}}_{\rm DP},
\label{eq:decay_plane_definition}
\end{align}
in which momenta are expressed in the \Lc rest frame.
With this definition the $\alpha$ Euler angle is the azimuthal angle of the proton in the \Lc polarisation frame, $\phi_p$, the $\beta$ angle is the polar angle of the proton, $\theta_p$, and the $\gamma$ angle is the signed angle between the plane formed by the proton and the \Lc quantisation axis and the plane formed by the kaon and the pion, named $\chi$.

The five variables describing a uniform phase space density can be chosen to be
\begin{equation}
\Omega = (\mqpk,\mqkpi,\cos\theta_p,\phi_p,\chi).
\end{equation}

To simplify the \Lcpkpi amplitude model expression, it is useful to separate invariant mass and decay orientation degrees of freedom applying the Dalitz plot decomposition proposed in Ref.~\cite{Mikhasenko:2019rjf}. Moreover, the properties following from rotational invariance described in Section~\ref{sec:properties_polarised_decay_rate} are enforced by construction. For the \Lcpkpi decay the decomposition is written as
\begin{equation}
\mathcal{A}_{m_{\Lc},m_p}(\Omega) = \sum_{\nu_{\Lc}} D^{*1/2}_{m_{\Lc},\nu_{\Lc}}(\phi_p,\theta_p,\chi) \text{ } \mathcal{O}_{\nu_{\Lc},m_p}(\mqpk,\mqkpi),
\label{eq:DPD_decomposition}
\end{equation}
in which the Wigner $D$ matrix describes the rotation of $|1/2,\mu_{\Lc}\rangle$ \Lc spin states to those associated to the decay plane system Eq.~\eqref{eq:decay_plane_definition}, $|1/2,\nu_{\Lc}\rangle$. The term $\mathcal{O}_{\nu_{\Lc},m_p}(\mqpk,\mqkpi)$ is the decay amplitude in terms of $|1/2,\nu_{\Lc}\rangle$ states and proton states defined in the canonical spin system reached from the \Lc decay plane system, $\Ket{1/2,m_p}$. These proton states are needed for the matching of proton spin states among different decay chains.

\subsection{General Study of the \Lcpkpi Polarised Decay Rate}
\label{sec:general_study_pol}
In the following we present the general structure of the decay rate written applying the Dalitz plot decomposition for different polarisation characteristics. We consider the decay rate decomposed into unpolarised, longitudinal and orthogonal polarisation parts Eqs.~\eqref{eq:unpolarised_decay_rate},~\eqref{eq:longitudinal_pol_decay_rate},~\eqref{eq:orthogonal_pol_decay_rate}, with Eq.~\eqref{eq:DPD_decomposition} applied to the decay amplitudes.

The unpolarised decay rate Eq.~\eqref{eq:unpolarised_decay_rate}is simply
\begin{equation}
p_{\rm unpol}(\mqpk,\mqkpi) = \sum_{\nu_{\Lc},m_p = \pm 1/2} \left| \mathcal{O}_{\nu_{\Lc},m_p} \right|^2,
\label{eq:Lcpkpi_decay_rate_unpol}
\end{equation}
because of the orthogonality of Wigner $D$-matrices
\begin{equation}
\sum_m D^{*1/2}_{m,\nu}(\phi_p,\theta_p,\chi) D^{1/2}_{m,\nu'}(\phi_p,\theta_p,\chi) \propto \delta_{\nu,\nu'},
\end{equation}
and does not depend on the orientation angles for rotational invariance.

The decay rate for longitudinal polarisation Eq.~\eqref{eq:longitudinal_pol_decay_rate} can be written in the form
\begin{align}
p_{\rm long}(\Omega,P_z) = P_z \left[\cos\theta_p \text{ } A(\mqpk,\mqkpi) -2\sin\theta_p \text{ } \re B(\mqpk,\mqkpi,\chi) \right],
\label{eq:Lcpkpi_decay_rate_long}
\end{align}
with
\begin{align}
A(\mqpk,\mqkpi) \equiv \sum_{m_p} \left( \left| \mathcal{O}_{1/2,m_p} \right|^2 - \left| \mathcal{O}_{-1/2,m_p} \right|^2 \right),\\
B(\mqpk,\mqkpi,\chi) \equiv \sum_{m_p} \left( \exp i\chi \text{ } \mathcal{O}_{1/2,m_p} \mathcal{O}^*_{-1/2,m_p}\right).
\end{align}
The first term introduces a simple $\cos\theta_p$ linear dependence; the behaviour of the second term can be seen as follows. Let's write
\begin{equation}
\mathcal{O}_{1/2,m_p} \mathcal{O}^*_{-1/2,m_p} \equiv P_{m_p} \exp i \Phi_{m_p},
\end{equation}
so that 
\begin{align}
\re B(\mqpk,\mqkpi,\chi) &= \sum_{m_p} P_{m_p} \cos \left( \chi + \Phi_{m_p} \right), \nonumber\\
&\equiv P \cos \left( \chi + \Phi \right)
\end{align}
is the sum of two cosine functions with different amplitude and phase but same frequency in $\chi$, which is equivalent to a single cosine function with some amplitude $P$ and phase $\Phi$. Since
\begin{equation}
\int_{-1}^{1} \sin\theta_p \text{ } d\cos\theta_p = \frac{\pi}{2},
\label{eq:sin_theta_p_integral}
\end{equation}
the oscillatory dependence of the $\re B$ term is visible in the $\chi$ projection\footnote{With projection we refer to the one-dimensional decay distribution obtained integrating over all the phase space variables but one.} of the amplitude model. Instead, the $\sin\theta_p$ dependence is not visible in the $\cos\theta_p$ projection since
\begin{equation}
\int_{-\pi}^{\pi} P \cos \left( \chi + \Phi \right) d\chi = 0.
\end{equation}

The decay rate for orthogonal polarisation can be written in the form
\begin{align}
p_{\rm orth}(\Omega,P_x,P_y) &= (P_x \cos\phi_p - P_y \sin\phi_p) \left[ \sin\theta_p A(\mqpk,\mqkpi) \right. \nonumber\\
&+ \left. 2\cos\theta_p \re B(\mqpk,\mqkpi,\chi) + 2i \im B(\mqpk,\mqkpi,\chi) \right],
\end{align}
by exploiting the relation
\begin{equation}
a\cos^2\alpha - a^*\sin^2\alpha = \re a \cos 2\alpha + i \im a.
\end{equation}
For Eq.~\eqref{eq:sin_theta_p_integral}, the cosine (minus sine) dependence shows up in the $\phi_p$ projection, providing a clear signature for the presence of $P_x$ ($P_y$) orthogonal polarisation. Instead, orthogonal polarisation does not introduce effects in $\theta_p$ and $\chi$ projections, since
\begin{equation}
\int_{-\pi}^{\pi} \cos\phi_p d\phi_p = \int_{-\pi}^{\pi} \sin\phi_p d\phi_p = 0.
\end{equation}

The predicted dependence of the phase space variable projections on different \Lc polarisation components is shown with the use of Monte Carlo generated data for the toy amplitude model defined in Section~\ref{sec:toy_fit} in Appendix~\ref{sec:projections}.

\subsection{Decay Amplitudes in the Helicity Formalism}
\label{sec:decay_amplitudes}
The \Lcpkpi decay amplitudes $\mathcal{O}_{\nu_{\Lc},m_p}$ are written in the helicity formalism described in Section~\ref{sec:helicity_formalism}.
We first consider the decay chain $\Lc\to pK^*(\to K^-\pi^+)$. The weak decay $\Lc\to pK^*$ can be described by Eq.~\eqref{eq:two_body_amplitude_corrected} applied starting from the decay plane coordinate system,
\begin{equation}
\mathcal{A}^{\Lc\to pK^*}_{\nu_{\Lc},m_p,\bar{\lambda}_{K^*}} = \mathcal{H}^{\Lc\to pK^*}_{m_p,\bar{\lambda}_{K^*}} \delta_{\nu_{\Lc},m_p+\bar{\lambda}_{K^*}},
\end{equation}
so that the amplitude is written in terms of the proton spin $m_p$ and the $K^*$ opposite helicity $\bar{\lambda}_{K^*}$. Since no rotation of spin states is involved, the $D$-matrix becomes a constraint on the helicity values $m_p+\bar{\lambda}_{K^*}=\nu_{\Lc}$.

For spin zero $K^*$ resonances the angular momentum conservation relations Eq.~\eqref{eq:allowed_helicity_couplings} allow two complex couplings corresponding to $m_p=\pm 1/2$; for higher spin resonances four couplings are allowed, corresponding to $\lbrace m_p=1/2$, $\bar{\lambda}_{K^*}=0,-1\rbrace$ and $\lbrace m_p=-1/2$, $\bar{\lambda}_{K^*}=0,1 \rbrace$. The couplings are independent of each other because of parity violation in weak decays. The strong decay $K^*\to K^-\pi^+$ contribution is
\begin{equation}
\mathcal{A}^{K^*\to K^-\pi^+}_{\bar{\lambda}_{K^*}} = \mathcal{H}^{K^*\to K^-\pi^+}_{0,0} d^{*S_{K^*}}_{\bar{\lambda}_{K^*},0}(\bar{\theta}_K)\mathcal{R}(\mqkpi),
\label{eq:amplitude_Kstar_Kpi_DPD}
\end{equation}
in which $\mathcal{R}(\mqkpi)$ is the lineshape of the $K^*$ resonance and $\bar{\theta}_{K}$ is the kaon momentum signed polar angle in the $K^*$ opposite-helicity coordinate system,
\begin{equation}
\bar{\theta}_{K} = \mathrm{atan2}\left(p_x^{K^*}(K^-),p_z^{K^*}(K^-)\right).
\end{equation}
Signed polar angles are used as helicity angles in order to have rotations only around the $y$ axis of the decay plane system\footnote{Otherwise, the use of positive polar angles would require additional azimuthal rotations around the $z$ axis (to flip the $y$ axis direction) complicating unnecessarily the expression of the helicity amplitudes.}.
In the fit model the coupling $\mathcal{H}^{K^*\to K^-\pi^+}_{0,0}$ can not be determined independently of $\mathcal{H}^{\Lc\to K^* p}_{m_p,\bar{\lambda}_{K^*}}$ couplings, therefore it is set equal to 1 and absorbed into the latter.

Considering the decay chain $\Lc\to\Lz^*(\to pK^-)\pi^+$, the weak decay $\Lc\to\Lz^*\pi^+$ is described by Eq.~\eqref{eq:two_body_amplitude_corrected} as
\begin{equation}
\mathcal{A}^{\Lc\to\Lz^*\pi^+}_{\nu_{\Lc},\lambda_{\Lz^*}} = \mathcal{H}^{\Lc\to\Lz^*\pi^+}_{\lambda_{\Lz^*},0} d^{1/2}_{\nu_{\Lc},\lambda_{\Lz^*}}(\theta_{\Lz^*}),
\end{equation}
in which $\lambda_{\Lz^*}$ is the $\Lz^*$ helicity system reached from the \Lc system and $\theta_{\Lz^*}$ is the signed polar angle of the $\Lz^*$ momentum, defined as
\begin{equation}
\theta_{\Lz^*} = \mathrm{atan2}\left(p_x^{\Lc}(\Lz^*),p_z^{\Lc}(\Lz^*)\right).
\end{equation}
The angular momentum conservation relations Eq.~\eqref{eq:allowed_helicity_couplings} allow two helicity couplings, $\lambda_{\Lz^*}=\pm 1/2$, to fit for each resonance whatever $J_{\Lz^*}$ is.
The strong decay $\Lz^*\to pK^-$ is described by
\begin{equation}
\mathcal{A}^{\Lz^*\to pK^-}_{\lambda_{\Lz^*},\lambda^{\Lz^*}_p} = \mathcal{H}^{\Lz^*\to pK^-}_{\lambda^{\Lz^*}_p,0} d^{S_{\Lz^*}}_{\lambda_{\Lz^*},\lambda^{\Lz^*}_p}(\theta^{\Lz^*}_p)\mathcal{R}(\mqpk),
\end{equation}
in which $\lambda^{\Lz^*}_p$ is the proton helicity, $\theta^{\Lz^*}_p$ the proton signed polar angle in the helicity coordinate system reached from the $\Lz^*$ resonance. Since strong decays conserve parity the two helicity couplings corresponding to $\lambda^{\Lz^*}_p=\pm 1/2$ are related,
\begin{equation}
\mathcal{H}^{\Lz^*\to pK^-}_{-\lambda^{\Lz^*}_p,0} = -P_{\Lz^*} (-1)^{S_{\Lz^*}-1/2} \mathcal{H}^{\Lz^*\to pK^-}_{\lambda^{\Lz^*}_p,0},
\end{equation}
in which $P_{\Lz^*}$ is the parity of the $\Lz^*$ resonance and the proton and kaon parities $P_p=1$, $P_K=-1$ have been inserted. In the fit model these couplings are absorbed into $\mathcal{H}^{\Lc\to\Lz^*\pi^+}_{\lambda_{\Lz^*},0}$, setting $\mathcal{H}^{\Lz^*\to pK^-}_{+1/2,0}=1$ and $\mathcal{H}^{\Lz^*\to pK^-}_{-1/2,0}=-P_{\Lz^*} (-1)^{S_{\Lz^*}-1/2}$.

The matching of proton spin states from the $\Lz^*$ helicity system to the canonical system it is performed applying the method of Ref.~\cite{Marangotto:2019ucc} to the case of the Dalitz-plot decomposition. Indeed, the transformation sequence applied to reach the proton helicity frame must be ``undone'' step-by-step in order to ensure a consistent phase definition of fermion spin states. Three rotations must be applied to the proton spin system: two of angles $\theta^{\Lz^*}_p$ and $\theta_{\Lz^*}$, plus the Wigner rotation accounting for the different boost sequence applied to reach the two systems. The Wigner rotation can be written in angle-axis decomposition~\cite{Gourgoulhon}, with angle
\begin{equation}
\alpha^W_{\Lz^*} = \arccos\left[ \frac{\left( 1+ \gamma^{\Lc}_p + \gamma^{\Lc}_{\Lz^*} + \gamma^{\Lz^*}_p \right)^2}{(1+\gamma^{\Lc}_p) (1+\gamma^{\Lc}_{\Lz^*}) (1+\gamma^{\Lz^*}_p)} -1 \right],
\end{equation}
with $\gamma^A_B$ the gamma factor of the boost connecting $A$ and $B$ systems,
and axis
\begin{equation}
\bm{a}^W_{\Lz^*} = \frac{\bm{p}^{\Lc}(\Lz^*) \times \bm{p}^{\Lz^*}(p)}{\left| \bm{p}^{\Lc}(\Lz^*) \times \bm{p}^{\Lz^*}(p) \right|} = \bm{\hat{y}}_{\rm DP},
\end{equation}
\ie the $y$ axis of the decay plane coordinate system. All these rotations are around the same $y$ axis, combined into one rotation $R_y(\beta_{\Lz^*})$, with
\begin{equation}
\beta_{\Lz^*} = \theta^{\Lz^*}_p+\theta_{\Lz^*}+\alpha^W_{\Lz^*}.
\end{equation}

Considering the third decay chain $\Lc\to\Deltares^{++*}(\to p\pi^+)K^-$, the weak decay $\Lc\to\Deltares^{++*}K^-$ is described by
\begin{equation}
\mathcal{A}^{\Lc\to\Deltares^{++*}K^-}_{\nu_{\Lc},\lambda_{\Deltares^*}} = \mathcal{H}^{\Lc\to\Deltares^{++*}K^-}_{\lambda_{\Deltares^*},0} d^{1/2}_{\nu_{\Lc},\lambda_{\Deltares^*}}(\theta_{\Deltares^*}),
\end{equation}
in which $\lambda_{\Deltares^*}$ is the $\Deltares^*$ helicity and $\theta_{\Deltares^*}$ is the signed polar angle of the $\Deltares^*$ momentum in the \Lc rest frame with decay plane coordinate system. As for the $\Lz^*$ decay chain, there are two helicity couplings corresponding to $\lambda_{\Deltares^*}=\pm 1/2$ to fit for each resonance. The strong decay $\Deltares^{++*}\to p\pi^+$ amplitude is written as
\begin{equation}
\mathcal{A}^{\Deltares^{++*}\to p\pi^+}_{\lambda_{\Deltares^*},\lambda^{\Deltares^*}_p} = \mathcal{H}^{\Deltares^{++*}\to p\pi^+}_{\lambda^{\Deltares^*}_p,0} d^{S_{\Deltares^*}}_{\lambda_{\Deltares^*},\lambda^{\Deltares^*}_p}(\theta^{\Deltares^*}_p)\mathcal{R}(\mqppi),
\end{equation}
in which $\lambda^{\Deltares^*}_p$ is the proton helicity and $\theta^{\Deltares^*}_p$ the signed polar angle defined in the $\Deltares^*$ helicity coordinate system. In the fit model the strong decay helicity couplings are absorbed into $\mathcal{H}^{\Lc\to\Deltares^{++*}K^-}_{\lambda_{\Deltares^*},0}$ setting them to $\mathcal{H}^{\Deltares^{++*}\to p\pi^+}_{+1/2,0}=1$ and $\mathcal{H}^{\Deltares^{++*}\to p\pi^+}_{-1/2,0}=-P_{\Deltares^*} (-1)^{S_{\Deltares^*}-1/2}$.

The matching of proton spin states from the $\Deltares^*$ helicity system to the canonical system is performed similarly to the $\Lz^*$ decay chain. The Wigner rotation angle is
\begin{equation}
\alpha^W_{\Deltares^*} = \arccos\left[ \frac{\left( 1+ \gamma^{\Lc}_p + \gamma^{\Lc}_{\Deltares^*} + \gamma^{\Deltares^*}_p \right)^2}{(1+\gamma^{\Lc}_p) (1+\gamma^{\Lc}_{\Deltares^*}) (1+\gamma^{\Deltares^*}_p)} -1 \right],
\end{equation}
around the axis
\begin{equation}
\bm{a}^W_{\Deltares^*} = \frac{\bm{p}^{\Lc}(\Deltares^*) \times \bm{p}^{\Deltares^*}(p)}{\left| \bm{p}^{\Lc}(\Deltares^*) \times \bm{p}^{\Deltares^*}(p) \right|} = -\bm{\hat{y}}_{\rm DP},
\end{equation}
which is opposite to the $y$ axis of the decay plane coordinate system. Therefore, the proton spin rotation can be written as $R_y(\beta_{\Deltares^*})$, with reversed Wigner angle sign
\begin{equation}
\beta_{\Deltares^*} = \theta^{\Deltares^*}_p+\theta_{\Deltares^*}-\alpha^W_{\Deltares^*}.
\end{equation}

The decay amplitudes for each decay chain are the product of two two-body decay amplitudes, summed over the proton helicities for $\Lz^*$ and $\Deltares^*$ chains,
\begin{align}
\mathcal{A}^{K^*}_{\nu_{\Lc},m_p,\bar{\lambda}_{K^*}} &= \mathcal{H}^{K^*}_{m_p,\bar{\lambda}_{K^*}} \delta_{\nu_{\Lc},m_p+\bar{\lambda}_{K^*}} d^{*S_{K^*}}_{\bar{\lambda}_{K^*},0}(\bar{\theta}_K)\mathcal{R}_{K^*}(\mqkpi),\\
\mathcal{A}^{\Lz^*}_{\nu_{\Lc},\lambda_{\Lz^*},m_p} &= \sum_{\lambda^{\Lz^*}_p} \mathcal{H}^{\Lz^*}_{\lambda_{\Lz^*},0} d^{1/2}_{\nu_{\Lc},\lambda_{\Lz^*}}(\theta_{\Lz^*}) d^{S_{\Lz^*}}_{\lambda_{\Lz^*},\lambda^{\Lz^*}_p}(\theta^{\Lz^*}_p)d^{1/2}_{m_p,\lambda^{\Lz^*}_p}(\beta_{\Lz^*})\mathcal{R}_{\Lz^*}(\mqpk),\label{eq:Lstar_decay_amplitude}\\
\mathcal{A}^{\Deltares^{++*}}_{\nu_{\Lc},\lambda_{\Deltares^*},m_p} &= \sum_{\lambda^{\Deltares^*}_p} \mathcal{H}^{\Deltares^{++*}}_{\lambda_{\Deltares^*},0} d^{1/2}_{\nu_{\Lc},\lambda_{\Deltares^*}}(\theta_{\Deltares^*})
d^{S_{\Deltares^*}}_{\lambda_{\Deltares^*},\lambda^{\Deltares^*}_p}(\theta^{\Deltares^*}_p)d^{1/2}_{m_p,\lambda^{\Deltares^*}_p}(\beta_{\Deltares^*})\mathcal{R}_{\Deltares^*}(\mqppi).\label{eq:Dstar_decay_amplitude}
\end{align}

The complete amplitude for the \Lcpkpi decay is obtained summing the amplitudes for all the intermediate resonances and their allowed helicity states,
\begin{align}
\mathcal{O}_{\nu_{\Lc},m_p}(\mqpk,\mqkpi) &= \sum_{i=1}^{N_{K^*}} \sum_{\bar{\lambda}_{K^*}} \mathcal{A}^{\Lc\to K^*_i(\to K^-\pi^+)p}_{\nu_{\Lc},m_p,\bar{\lambda}_{K^*}} \nonumber\\
&+ \sum_{j=1}^{N_{\Lz^*}} \sum_{\lambda_{\Lz^*}} \mathcal{A}^{\Lc\to\Lz_i^*(\to pK^-)\pi^+}_{\nu_{\Lc},\lambda_{\Lz^*},m_p} \nonumber\\
&+ \sum_{k=1}^{N_{\Deltares^*}} \sum_{\lambda_{\Deltares^{++*}}} \mathcal{A}^{\Lc\to\Deltares_k^{++*}K^-}_{\nu_{\Lc},\lambda_{\Deltares^*},m_p}.
\label{eq:helicity_amplitudes_Lc}
\end{align}

\subsection{Study of the Decay Rate}
\label{sec:decay_rate_study}
We divide the study of the decay rate into different parts, investigating which information each gives on the parameters describing the decay distribution. In particular, we will focus on helicity couplings and the polarisation degree. Each complex helicity coupling is equivalent to two real parameters, its real and imaginary part or its modulus and phase. Therefore, we have two real unknowns for each coupling plus the polarisation modulus. Each part of the decay rate having a functional form distinguishable from the others via an amplitude fit, yields a constraint on the combination of parameters involved in that contribution.

First, we study the part of the decay rate without interference terms, consisting of the sum of the rates associated to each single contribution. For the \Lcpkpi decay, each contribution is given by Eq.~\eqref{eq:decay_rate_3body_weak_strong_decay}, since intermediate resonances decay via parity-conserving strong interaction,
\begin{align}
p^{\rm non-int}(\Omega,P_z) &= \sum_{i=1}^{N_{K^*}} \left( F_{K_i^*} + P_z \alpha_{K_i^*}\cos\theta_{K^*} \right) |g^{S_{K^*_i}}_{K^*_i}(\mqkpi,\theta_K)|^2 \nonumber\\
&+ \sum_{j=1}^{N_{\Lz^*}} \left( F_{\Lz_j^*} + P_z \alpha_{\Lz_j^*} \cos\theta_{\Lz^*} \right) |g^{S_{\Lz^*_j}}_{\Lz^*_j}(\mqpk,\theta^{\Lz^*}_p)|^2 \nonumber\\
&+ \sum_{k=1}^{N_{\Deltares^*}} \left( F_{\Deltares_k^*} + P_z \alpha_{\Deltares_k^*}\cos\theta_{\Deltares^*} \right) |g^{S_{\Deltares^*_k}}_{\Deltares^*_j}(\mqppi,\theta^{\Deltares^*}_p)|^2,
\label{eq:Lcpkpi_decay_rate_non_int}
\end{align}
in which angles are defined in analogy with Eq.~\eqref{eq:decay_rate_3body_weak_strong_decay} and the $F_{R}$ values are the sum of the helicity couplings squared moduli for the resonance $R$,
\begin{equation}
F_{R} \equiv \sum_{\lambda_1,\bar{\lambda}_2} \left| \mathcal{H}^R_{\lambda_1,\bar{\lambda}_2} \right|^2.
\end{equation}
These are related to the resonance fit fractions $\mathcal{F}_{R}$ via
\begin{equation}
\mathcal{F}_{R} = \int F_{R} |g^{S_{R}}_{R}(m^2_R,\theta^R_1)|^2 d\Omega.
\end{equation}
The overall normalisation of the decay rate, ensuring it has unit integral over the phase space, is intended to be implicit in the definition of the helicity couplings.

The decay rate part without interference terms gives information on $F_{R}$ and the products $P_z \alpha_R$, but, again, it does not allow to separate the polarisation modulus from the $\alpha$ values without an independent measurement of at least one of them.

For later convenience, let's evaluate the non-interfering decay rate part associated to a $\Lz^*$ resonance in the Dalitz plot amplitude decomposition. Starting from Eq.~\eqref{eq:Lstar_decay_amplitude}, its expression greatly simplifies exploiting the orthogonality of $d$-matrices,
\begin{align}
\sum_{m_p} d^{1/2}_{m_p,\lambda^{\Lz^*}_p}(\beta_{\Lz^*}) d^{1/2}_{m_p,\lambda'^{\Lz^*}_p}(\beta_{\Lz^*}) &= \delta_{\lambda^{\Lz^*}_p,\lambda'^{\Lz^*}_p},\nonumber\\
\sum_{\lambda^{\Lz^*}_p} d^{S_{\Lz^*}}_{\lambda_{\Lz^*},\lambda^{\Lz^*}_p}(\theta^{\Lz^*}_p) d^{S_{\Lz^*}}_{\lambda'_{\Lz^*},\lambda^{\Lz^*}_p}(\theta^{\Lz^*}_p) &= \delta_{\lambda_{\Lz^*},\lambda'_{\Lz^*}} f(S_{\Lz^*},\theta^{\Lz^*}_p)^2,
\end{align}
obtaining
\begin{align}
\sum_{m_p}\mathcal{O}^{\Lz^*}_{\nu_{\Lc},m_p} \mathcal{O}^{*\Lz^*}_{\nu'_{\Lc},m_p} = \sum_{\lambda_{\Lz^*}} \left|\mathcal{H}^{\Lz^*}_{\lambda_{\Lz^*},0}\right|^2 \text{ } d^{1/2}_{\nu_{\Lc},\lambda_{\Lz^*}}(\theta_{\Lz^*})d^{1/2}_{\nu'_{\Lc},\lambda_{\Lz^*}}(\theta_{\Lz^*})|g^{S_{\Lz^*}}_{\Lz^*}(\mqpk,\theta^{\Lz^*}_p)|^2.
\end{align}
Explicitly we have
\begin{align}
\sum_{m_p}\left|\mathcal{O}^{\Lz^*}_{1/2,m_p}\right|^2 &= \left( \left|\mathcal{H}^{\Lz^*}_{1/2,0}\right|^2 \cos^2\frac{\theta_{\Lz^*}}{2} + \left|\mathcal{H}^{\Lz^*}_{-1/2,0}\right|^2 \sin^2\frac{\theta_{\Lz^*}}{2}\right) |g^{S_{\Lz^*}}_{\Lz^*}(\mqpk,\theta^{\Lz^*}_p)|^2, \nonumber\\
\sum_{m_p}\mathcal{O}^{\Lz^*}_{1/2,m_p} \mathcal{O}^{*\Lz^*}_{-1/2,m_p} &= \left( \left|\mathcal{H}^{\Lz^*}_{1/2,0}\right|^2 - \left|\mathcal{H}^{\Lz^*}_{-1/2,0}\right|^2 \right) \sin\frac{\theta_{\Lz^*}}{2}\cos\frac{\theta_{\Lz^*}}{2} \nonumber\\
&\times |g^{S_{\Lz^*}}_{\Lz^*}(\mqpk,\theta^{\Lz^*}_p)|^2 = \sum_{m_p}\mathcal{O}^{\Lz^*}_{-1/2,m_p} \mathcal{O}^{*\Lz^*}_{1/2,m_p}, \nonumber\\
\sum_{m_p} \left|\mathcal{O}^{\Lz^*}_{-1/2,m_p} \right|^2 &= \left( \left|\mathcal{H}^{\Lz^*}_{1/2,0}\right|^2 \sin^2\frac{\theta_{\Lz^*}}{2} + \left|\mathcal{H}^{*\Lz^*}_{-1/2,0}\right|^2 \cos^2\frac{\theta_{\Lz^*}}{2} \right) |g^{S_{\Lz^*}}_{\Lz^*}(\mqpk,\theta^{\Lz^*}_p)|^2.
\end{align}
The contribution to the unpolarised decay rate Eq.~\eqref{eq:Lcpkpi_decay_rate_unpol} is
\begin{align}
p_{\rm unpol}(\mqpk,\mqkpi) &= \sum_{\nu_{\Lc},m_p} \left|\mathcal{O}^{\Lz^*}_{\nu_{\Lc},m_p}\right|^2 = F_{\Lz^*} |g^{S_{\Lz^*}}_{\Lz^*}(\mqpk,\theta^{\Lz^*}_p)|^2,
\label{eq:Lcpkpi_decay_rate_non_int_unpol}
\end{align}
while the contribution to the longitudinal polarisation rate Eq.~\eqref{eq:Lcpkpi_decay_rate_long} is
\begin{align}
p_{\rm long}(\Omega,P_z) &= P_z \left[ \cos\theta_p \text{ } \sum_{m_p} \left( \left|\mathcal{O}^{\Lz^*}_{1/2,m_p}\right|^2 - \left|\mathcal{O}^{\Lz^*}_{-1/2,m_p}\right|^2  \right)\right.\nonumber\\
& \left. \hspace{8pt} -2\sin\theta_p \text{ } \re \sum_{m_p} \left( \exp i\chi \text{ }\mathcal{O}^{\Lz^*}_{1/2,m_p} \mathcal{O}^{*\Lz^*}_{-1/2,m_p}\right)\right]\nonumber\\
&= P_z \alpha_{\Lz^*} \left(\cos\theta_p \cos\theta_{\Lz^*} - \sin\theta_p \sin\theta_{\Lz^*} \cos\chi \right) |g^{S_{\Lz^*}}_{\Lz^*}(\mqpk,\theta^{\Lz^*}_p)|^2.
\label{eq:Lcpkpi_decay_rate_non_int_long}
\end{align}
The above expressions are equivalent to Eq.~\eqref{eq:Lcpkpi_decay_rate_non_int}, the only difference being in the choice of the angles describing the decay distributions. Indeed, the definition of the helicity angle $\theta_{\Lz^*}$ is different in the two cases: while in Eqs.~\eqref{eq:Lcpkpi_decay_rate_non_int_unpol},~\eqref{eq:Lcpkpi_decay_rate_non_int_long} it is independent of the decay orientation angles ($\theta_p$, $\phi_p$, $\chi$), in Eq.~\eqref{eq:Lcpkpi_decay_rate_non_int} $\theta_{\Lz^*}$ depends on a combination of invariant mass and decay orientation degrees of freedom. Of course, nothing changes in terms of sensitivity to the decay parameters: the $\chi$ angle dependent term does not add information to the \Lcpkpi decay parameters.

The $\Lz^*$ contribution is characterised by 4 unknowns related to couplings plus the polarisation modulus, which is shared with the other contributions. This part of the decay rate places two constraints determining the sum of the helicity coupling moduli and their relative difference up to the polarisation factor.

Now, we consider the interference terms between resonance contributions, showing in which way they allow to get maximum information on the decay amplitudes, constraining all the parameters describing the \Lcpkpi decay distributions. 
There are two kind of interference terms: those arising from resonances belonging to the same decay chain and those coming from resonances associated to different decay chains. In the following we will consider one example for each type.

\subsubsection{Interference Term $\Lz^*-\Lz'^*$}
\label{sec:Lambda_Lambda_int}
Let's consider the interference term between two $\Lz^*$ resonances. Its computation is analogous to the $\Lz^*$ decay rate, since both resonances belong to the same decay chain, sharing the same rotation angles,
\begin{align}
\sum_{m_p}\mathcal{O}^{\Lz^*}_{\nu_{\Lc},m_p} \mathcal{O}^{*\Lz'^*}_{\nu'_{\Lc},m_p} = &\sum_{\lambda_{\Lz^*}} \mathcal{H}^{\Lz^*}_{\lambda_{\Lz^*},0} \mathcal{H}^{*\Lz'^*}_{\lambda_{\Lz^*},0} \text{ } d^{1/2}_{\nu_{\Lc},\lambda_{\Lz^*}}(\theta_{\Lz^*})d^{1/2}_{\nu'_{\Lc},\lambda_{\Lz^*}}(\theta_{\Lz^*}) G^{S_{\Lz^*},S_{\Lz'^*}}_{\Lz^*,\Lz'^*}(\mqpk,\theta^{\Lz^*}_p),
\end{align}
in which
\begin{equation}
G^{S_{\Lz^*},S_{\Lz'^*}}_{\Lz^*,\Lz'^*}(\mqpk,\theta^{\Lz^*}_p) \equiv f(S_{\Lz^*},\theta^{\Lz^*}_p) f(S_{\Lz'^*},\theta^{\Lz^*}_p) \mathcal{R}_{\Lz^*}(\mqpk)\mathcal{R}_{\Lz'^*}(\mqpk).
\end{equation}
Explicitly (leaving out $G$ function arguments),
\begin{align}
\sum_{m_p}\mathcal{O}^{\Lz^*}_{1/2,m_p} \mathcal{O}^{*\Lz'^*}_{1/2,m_p} &= \left( \mathcal{H}^{\Lz^*}_{1/2,0} \mathcal{H}^{*\Lz'^*}_{1/2,0}\cos^2\frac{\theta_{\Lz^*}}{2} + \mathcal{H}^{\Lz^*}_{-1/2,0} \mathcal{H}^{*\Lz'^*}_{-1/2,0}\sin^2\frac{\theta_{\Lz^*}}{2}\right) G,\nonumber\\
\sum_{m_p}\mathcal{O}^{\Lz^*}_{1/2,m_p} \mathcal{O}^{*\Lz'^*}_{-1/2,m_p} &= \left( \mathcal{H}^{\Lz^*}_{1/2,0}\mathcal{H}^{*\Lz'^*}_{1/2,0} - \mathcal{H}^{\Lz^*}_{-1/2,0} \mathcal{H}^{*\Lz'^*}_{-1/2,0} \right) \sin\frac{\theta_{\Lz^*}}{2}\cos\frac{\theta_{\Lz^*}}{2} G \nonumber\\
&= \sum_{m_p}\mathcal{O}^{\Lz^*}_{-1/2,m_p} \mathcal{O}^{*\Lz'^*}_{1/2,m_p}, \nonumber\\
\sum_{m_p}\mathcal{O}^{\Lz^*}_{-1/2,m_p} \mathcal{O}^{*\Lz'^*}_{-1/2,m_p} &= \left( \mathcal{H}^{\Lz^*}_{1/2,0} \mathcal{H}^{*\Lz'^*}_{1/2,0}\sin^2\frac{\theta_{\Lz^*}}{2} + \mathcal{H}^{\Lz^*}_{-1/2,0} \mathcal{H}^{*\Lz'^*}_{-1/2,0}\cos^2\frac{\theta_{\Lz^*}}{2} \right) G.
\end{align}
Let's consider the contribution of this interference term to the decay rate, including its complex conjugate corresponding to the exchange $\Lz^* \leftrightarrow \Lz'^*$. The unpolarised decay rate part Eq.~\eqref{eq:Lcpkpi_decay_rate_unpol} is
\begin{align}
p_{\rm unpol}(\mqpk,\mqkpi) &= 2\re\sum_{\nu_{\Lc},m_p} \mathcal{O}^{\Lz^*}_{\nu_{\Lc},m_p} \mathcal{O}^{*\Lz'^*}_{\nu'_{\Lc},m_p} \nonumber\\
&= 2\re \left( F_{\Lz^*\Lz'^*} G \right),
\end{align}
while the longitudinal polarisation part Eq.~\eqref{eq:Lcpkpi_decay_rate_long} is
\begin{align}
p_{\rm long}(\Omega,P_z) &= P_z \left\lbrace \cos\theta_p \text{ } 2\re \sum_{m_p} \left( \mathcal{O}^{\Lz^*}_{1/2,m_p} \mathcal{O}^{*\Lz'^*}_{1/2,m_p} - \mathcal{O}^{\Lz^*}_{-1/2,m_p} \mathcal{O}^{*\Lz'^*}_{-1/2,m_p} \right)\right.\nonumber\\
& \left. \hspace{8pt} -2\sin\theta_p \text{ } \re \sum_{m_p} \left[ \exp i\chi \text{ } \left(\mathcal{O}^{\Lz^*}_{1/2,m_p} \mathcal{O}^{*\Lz'^*}_{-1/2,m_p} + \mathcal{O}^{\Lz^*}_{-1/2,m_p} \mathcal{O}^{*\Lz'^*}_{1/2,m_p}\right)\right]\right\rbrace\nonumber\\
&= 2 P_z \left[\cos\theta_p \cos\theta_{\Lz^*} \text{ } \re \left(\alpha_{\Lz^*\Lz'^*} G\right) - \sin\theta_p \sin\theta_{\Lz^*} \text{ } \re \left( \exp i\chi \alpha_{\Lz^*\Lz'^*} G \right) \right],
\end{align}
which probe the combinations
\begin{align}
F_{\Lz^*\Lz'^*} \equiv \mathcal{H}^{\Lz^*}_{1/2,0} \mathcal{H}^{*\Lz'^*}_{1/2,0} + \mathcal{H}^{\Lz^*}_{-1/2,0} \mathcal{H}^{*\Lz'^*}_{-1/2,0},\nonumber\\
\alpha_{\Lz^*\Lz'^*} \equiv \mathcal{H}^{\Lz^*}_{1/2,0} \mathcal{H}^{*\Lz'^*}_{1/2,0} - \mathcal{H}^{\Lz^*}_{-1/2,0} \mathcal{H}^{*\Lz'^*}_{-1/2,0}.
\end{align}

The structure of this interference term is similar to the $\Lz^*$ decay rate: a $F_{\Lz^*\Lz'^*}$ term probed by the unpolarised decay rate and a polarisation-dependent term driven by a parity-violating asymmetry parameter $\alpha_{\Lz^*\Lz'^*}$.

To better study the constraints given by this interference term, let's write both complex couplings and the $G$ function in modulus-phase decomposition,
\begin{align}
\mathcal{H}^{\Lz^*}_{1/2,0} &= P^{\Lz^*}_{1/2,0} \exp i \Phi^{\Lz^*}_{1/2,0},\nonumber\\
G &= |G| \exp i \phi_G.
\end{align}
We have
\begin{align}
\re \left( F_{\Lz^*\Lz'^*} G \right)/|G| &= P^{\Lz^*}_{1/2,0} P^{\Lz'^*}_{1/2,0} \cos (\Phi^{\Lz^*}_{1/2,0} - \Phi^{\Lz'^*}_{1/2,0}+ \phi_G) \nonumber\\
&+ P^{\Lz^*}_{-1/2,0} P^{\Lz'^*}_{-1/2,0} \cos (\Phi^{\Lz^*}_{-1/2,0} - \Phi^{\Lz'^*}_{-1/2,0}+ \phi_G), \nonumber\\
P_z \re \left(\alpha_{\Lz^*\Lz'^*} G\right)/|G| &= P_z P^{\Lz^*}_{1/2,0} P^{\Lz'^*}_{1/2,0} \cos (\Phi^{\Lz^*}_{1/2,0} - \Phi^{\Lz'^*}_{1/2,0}+ \phi_G) \nonumber\\
&- P_z P^{\Lz^*}_{-1/2,0} P^{\Lz'^*}_{-1/2,0} \cos (\Phi^{\Lz^*}_{-1/2,0} - \Phi^{\Lz'^*}_{-1/2,0}+ \phi_G)
\end{align}
Since the phase $\phi_G$ has in general a non-trivial dependence on the resonance invariant mass (\eg relativistic Breit-Wigner lineshapes have such behaviour), the two conditions place four constraints on decay parameters: indeed one has
\begin{equation}
\cos (\Phi^{\Lz^*}_{1/2,0} - \Phi^{\Lz'^*}_{1/2,0}+ \phi_G) = \cos(\Phi^{\Lz^*}_{1/2,0} - \Phi^{\Lz'^*}_{1/2,0})\cos\phi_G - \sin(\Phi^{\Lz^*}_{1/2,0} - \Phi^{\Lz'^*}_{1/2,0})\sin\phi_G,
\label{eq:phase_constraint}
\end{equation}
showing that phase differences among couplings can be probed, not only their cosine, since $\cos\phi_G$ and $\sin\phi_G$ functional forms are separable by an amplitude fit. The constraint given by the $\chi$-dependent term is redundant.

Considering the decay rate part associated to $\Lz$ and $\Lz'^*$ resonances we have 9 unknowns (8 real couplings + $P_z$), and 8 constraints (4 from single decay rates + 4 from the interference term): again, we miss a condition to determine separately the set of helicity couplings and the polarisation.

What happens if we consider a set of three interfering $\Lz^*$ resonances? We have 13 unknowns and 16 constraints (now 12 coming from the three interference terms) and therefore the possibility to measure both the full set of helicity couplings and the polarisation. Practically, to have significant interference effects the three resonances must feature a significant overlap in invariant mass dependence (otherwise the lineshape product $\mathcal{R}_{\Lz^*}(\mqpk)\mathcal{R}_{\Lz'^*}(\mqpk)$ vanishes) which can make their separation in the amplitude fit difficult, especially if they have the same spin.

\subsubsection{Interference Term $\Lz^*-\Deltares^*$}
\label{sec:Lambda_Delta_int}
Let's consider the interference term between one $\Lz^*$ and one $\Deltares^*$ resonance. Starting from Eqs.~\eqref{eq:Lstar_decay_amplitude},~\eqref{eq:Dstar_decay_amplitude} its expression is
\begin{align}
\sum_{m_p}\mathcal{O}^{\Lz^*}_{\nu_{\Lc},m_p} \mathcal{O}^{*\Deltares^*}_{\nu'_{\Lc},m_p} &= \sum_{m_p}\sum_{\lambda_{\Lz^*}} \mathcal{H}^{\Lz^*}_{\lambda_{\Lz^*},0} d^{1/2}_{\nu_{\Lc},\lambda_{\Lz^*}}(\theta_{\Lz^*}) d^{S_{\Lz^*}}_{\lambda_{\Lz^*},\lambda^{\Lz^*}_p}(\theta^{\Lz^*}_p)d^{1/2}_{m_p,\lambda^{\Lz^*}_p}(\beta_{\Lz^*})
\nonumber\\
&\hspace{18pt}\times \sum_{\lambda_{\Deltares^*}} \mathcal{H}^{*\Deltares^*}_{\lambda_{\Deltares^*},0} d^{1/2}_{\nu'_{\Lc},\lambda_{\Deltares^*}}(\theta_{\Deltares^*})
d^{S_{\Deltares^*}}_{\lambda_{\Deltares^*},\lambda^{\Deltares^*}_p}(\theta^{\Deltares^*}_p)d^{1/2}_{m_p,\lambda^{\Deltares^*}_p}(\beta_{\Deltares^*})\nonumber\\
&\hspace{18pt}\times G^{S_{\Lz^*},S_{\Deltares^*}}_{\Lz^*,\Deltares^*}(\mqpk,\mqppi,\theta^{\Lz^*}_p,\theta^{\Deltares^*}_p),
\end{align}
with
\begin{equation}
G^{S_{\Lz^*},S_{\Deltares^*}}_{\Lz^*,\Deltares^*}(\mqpk,\mqppi,\theta^{\Lz^*}_p,\theta^{\Deltares^*}_p) \equiv f(S_{\Lz^*},\theta^{\Lz^*}_p) f(S_{\Deltares^*},\theta^{\Deltares^*}_p) \mathcal{R}_{\Lz^*}(\mqpk)\mathcal{R}_{\Deltares^*}(\mqppi).
\end{equation}
This interference term is more complicated than the previous one since the two resonances belong to different decay chains, characterised by different rotation angles. However, trigonometric functions arrange in such a way the proton rotation angles enter the decay rate only via the combination
\begin{equation}
\gamma \equiv \frac{\theta^{\Lz^*}_p-\theta^{\Deltares^*}_p-\beta_{\Lz^*}+\beta_{\Deltares^*}}{2}.
\end{equation}

Explicitly (leaving out function arguments),
\begin{align}
\sum_{m_p}\mathcal{O}^{\Lz^*}_{1/2,m_p} \mathcal{O}^{*\Deltares^*}_{1/2,m_p} &= \left[ \mathcal{H}^{\Lz^*}_{1/2,0} \mathcal{H}^{*\Deltares^*}_{1/2,0} \cos\frac{\theta_{\Lz^*}}{2}\cos\frac{\theta_{\Deltares^*}}{2} \cos\gamma \right. \nonumber\\
&+ \mathcal{H}^{\Lz^*}_{1/2,0} \mathcal{H}^{*\Deltares^*}_{-1/2,0} \cos\frac{\theta_{\Lz^*}}{2}\sin\frac{\theta_{\Deltares^*}}{2} \sin\gamma \nonumber\\
&- \mathcal{H}^{\Lz^*}_{-1/2,0} \mathcal{H}^{*\Deltares^*}_{1/2,0} \sin\frac{\theta_{\Lz^*}}{2}\cos\frac{\theta_{\Deltares^*}}{2} \sin\gamma \nonumber\\
&+ \left. \mathcal{H}^{\Lz^*}_{-1/2,0} \mathcal{H}^{*\Deltares^*}_{-1/2,0} \sin\frac{\theta_{\Lz^*}}{2}\sin\frac{\theta_{\Deltares^*}}{2} \cos\gamma \right] G,
\end{align}
\begin{align}
\sum_{m_p}\mathcal{O}^{\Lz^*}_{1/2,m_p} \mathcal{O}^{*\Deltares^*}_{-1/2,m_p} &= \left[ \mathcal{H}^{\Lz^*}_{1/2,0} \mathcal{H}^{*\Deltares^*}_{1/2,0} \cos\frac{\theta_{\Lz^*}}{2}\sin\frac{\theta_{\Deltares^*}}{2} \cos\gamma \right. \nonumber\\
&- \mathcal{H}^{\Lz^*}_{1/2,0} \mathcal{H}^{*\Deltares^*}_{-1/2,0} \cos\frac{\theta_{\Lz^*}}{2}\cos\frac{\theta_{\Deltares^*}}{2} \sin\gamma \nonumber\\
&- \mathcal{H}^{\Lz^*}_{-1/2,0} \mathcal{H}^{*\Deltares^*}_{1/2,0}\sin\frac{\theta_{\Lz^*}}{2}\sin\frac{\theta_{\Deltares^*}}{2} \sin\gamma \nonumber\\
&+ \left. \mathcal{H}^{\Lz^*}_{-1/2,0} \mathcal{H}^{*\Deltares^*}_{-1/2,0} \sin\frac{\theta_{\Lz^*}}{2}\cos\frac{\theta_{\Deltares^*}}{2} \cos\gamma \right] G,
\end{align}
\begin{align}
\sum_{m_p}\mathcal{O}^{\Lz^*}_{-1/2,m_p} \mathcal{O}^{*\Deltares^*}_{1/2,m_p} &= \left[ \mathcal{H}^{\Lz^*}_{1/2,0} \mathcal{H}^{*\Deltares^*}_{1/2,0} \sin\frac{\theta_{\Lz^*}}{2}\cos\frac{\theta_{\Deltares^*}}{2} \cos\gamma \right. \nonumber\\
&+ \mathcal{H}^{\Lz^*}_{1/2,0} \mathcal{H}^{*\Deltares^*}_{-1/2,0} \sin\frac{\theta_{\Lz^*}}{2}\sin\frac{\theta_{\Deltares^*}}{2} \sin\gamma \nonumber\\
&+ \mathcal{H}^{\Lz^*}_{-1/2,0} \mathcal{H}^{*\Deltares^*}_{1/2,0} \cos\frac{\theta_{\Lz^*}}{2}\cos\frac{\theta_{\Deltares^*}}{2} \sin\gamma \nonumber\\
&+ \left. \mathcal{H}^{\Lz^*}_{-1/2,0} \mathcal{H}^{*\Deltares^*}_{-1/2,0} \cos\frac{\theta_{\Lz^*}}{2}\sin\frac{\theta_{\Deltares^*}}{2} \cos\gamma \right] G,
\end{align}
\begin{align}
\sum_{m_p}\mathcal{O}^{\Lz^*}_{-1/2,m_p} \mathcal{O}^{*\Deltares^*}_{-1/2,m_p} &= \left[ \mathcal{H}^{\Lz^*}_{1/2,0} \mathcal{H}^{*\Deltares^*}_{1/2,0} \sin\frac{\theta_{\Lz^*}}{2}\sin\frac{\theta_{\Deltares^*}}{2} \cos\gamma \right. \nonumber\\
&- \mathcal{H}^{\Lz^*}_{1/2,0} \mathcal{H}^{*\Deltares^*}_{-1/2,0} \sin\frac{\theta_{\Lz^*}}{2}\cos\frac{\theta_{\Deltares^*}}{2} \sin\gamma \nonumber\\
&+ \mathcal{H}^{\Lz^*}_{-1/2,0} \mathcal{H}^{*\Deltares^*}_{1/2,0} \cos\frac{\theta_{\Lz^*}}{2}\sin\frac{\theta_{\Deltares^*}}{2} \sin\gamma \nonumber\\
&+ \left. \mathcal{H}^{\Lz^*}_{-1/2,0} \mathcal{H}^{*\Deltares^*}_{-1/2,0} \cos\frac{\theta_{\Lz^*}}{2}\cos\frac{\theta_{\Deltares^*}}{2} \cos\gamma \right] G.
\end{align}

The contribution to the unpolarised decay rate Eq.~\eqref{eq:Lcpkpi_decay_rate_unpol} is
\begin{align}
p_{\rm unpol}(\mqpk,\mqkpi) &= 2\re\sum_{\nu_{\Lc},m_p} \mathcal{O}^{\Lz^*}_{\nu_{\Lc},m_p} \mathcal{O}^{*\Deltares^*}_{\nu'_{\Lc},m_p} \nonumber\\
&= 2\re \left[ \left(\mathcal{H}^{\Lz^*}_{1/2,0} \mathcal{H}^{*\Deltares^*}_{1/2,0} + \mathcal{H}^{\Lz^*}_{-1/2,0}\mathcal{H}^{*\Deltares^*}_{-1/2,0}\right) \cos\left(\frac{\theta_{\Lz^*}-\theta_{\Deltares^*}}{2}\right) \cos\gamma \right. \nonumber\\
&\hspace{18pt}-\left.\left(\mathcal{H}^{\Lz^*}_{1/2,0}\mathcal{H}^{*\Deltares^*}_{-1/2,0} + \mathcal{H}^{\Lz^*}_{-1/2,0}\mathcal{H}^{*\Deltares^*}_{1/2,0}\right) \sin\left(\frac{\theta_{\Lz^*}-\theta_{\Deltares^*}}{2}\right) \sin\gamma \right] G,
\end{align}
and that to the $\chi$-independent longitudinal polarisation rate (the $\chi$-dependent term does not add information on the parameters) is
\begin{align}
p_{\rm long}(\Omega,P_z) &= P_z \cos\theta_p \text{ } 2\re \sum_{m_p} \left( \mathcal{O}^{\Lz^*}_{1/2,m_p} \mathcal{O}^{*\Deltares^*}_{1/2,m_p} - \mathcal{O}^{\Lz^*}_{-1/2,m_p} \mathcal{O}^{*\Deltares^*}_{-1/2,m_p} \right)\nonumber\\
&= P_z \cos\theta_p \text{ } 2\re \left[ \left( \mathcal{H}^{\Lz^*}_{1/2,0} \mathcal{H}^{*\Deltares^*}_{1/2,0} - \mathcal{H}^{\Lz^*}_{-1/2,0}\mathcal{H}^{*\Deltares^*}_{-1/2,0}\right) \cos\left(\frac{\theta_{\Lz^*}+\theta_{\Deltares^*}}{2}\right) \cos\gamma \right. \nonumber\\
&\hspace{48pt}+\left.\left(\mathcal{H}^{\Lz^*}_{1/2,0}\mathcal{H}^{*\Deltares^*}_{-1/2,0} -\mathcal{H}^{\Lz^*}_{-1/2,0}\mathcal{H}^{*\Deltares^*}_{1/2,0}\right) \sin\left(\frac{\theta_{\Lz^*}+\theta_{\Deltares^*}}{2}\right) \sin\gamma \right] G.
\end{align}

Following the discussion about the constraints placed by the $\Lz^*-\Lz'^*$ term, Section~\ref{sec:Lambda_Lambda_int}, each angular term determines the relative phase of each combination of helicity couplings, each corresponding to two constraints on real parameters, as from Eq.~\eqref{eq:phase_constraint}.
Considering the decay rate part associated to $\Lz$ and $\Deltares^*$ resonances we have 9 unknowns (8 real couplings + $P_z$), and 12 constraints (4 from single decay rates + 8 from the interference term); it is therefore possible to measure separately each parameter characterising the decay rate: real and imaginary parts of the complex couplings and the polarisation degree, if the latter is non-zero. Interference effects between different decay channels are more important than those in the same decay channel, the firsts giving more information on the amplitude model parameters.

In absence of polarisation, the constraints given by the unpolarised decay rate are insufficient to fully constrain helicity coupling values, as from the discussion in Section~\ref{sec:properties_polarised_decay_rate}. These results are numerically cross-checked by means of the toy amplitude fit presented in Section~\ref{sec:toy_fit}.

\subsubsection{Summary}
\label{sec:decay_rate_summary}
In the following we summarize the results of the study of the \Lcpkpi decay rate, focusing on which part of the decay rate gives information on which amplitude model parameter, Table~\ref{tab:summary_decay_rate}.

The sum of the helicity coupling moduli $F_R$ for each resonance (hence the fit fraction $\mathcal{F}_R$) can be measured from the decay rate without interference. The decay asymmetry parameters can be extracted from the same term but only in presence of a significant polarisation degree.

The sum of helicity coupling products $F_{RR'}$ characterising the interference between resonances belonging to the same decay channel can be measured in absence of polarisation, while the difference $\alpha_{RR'}$ needs non-zero polarisation.

The single helicity coupling complex values can be measured if their associate resonance has a significant interference with other two belonging to the same decay channel or one belonging to a different decay channel.

Assuming non-negligible parity-violation effects, the polarisation direction $\hat{\bm{P}}$ can be measured from the non-interfering decay rate, as following from the two-body decay rate for generic polarisation Eq.~\eqref{eq:two_body_decay_rate_generic_pol}. Instead, the polarisation modulus $|\bm{P}|$ can be determined separately from the helicity couplings only in presence of significant interference among three resonances in the same decay channel or two belonging to different decay channels.

According to the ongoing \Lcpkpi amplitude analysis from semileptonic decays at the LHCb experiment~\cite{Marangotto:2713231}, all the requirements needed for extracting maximum information from \Lcpkpi decay distributions are met: non-zero \Lc polarisation produced in the parent beauty hadron weak decay, significant parity-violation and resonant contributions in all its three decay channels with sizeable interference effects.

\begin{table}
\centering
\begin{tabular}{lll}
\toprule
Quantity & Decay rate part & Requirements\\
\midrule
$F_R$, $\mathcal{F}_R$ & Non-interfering & \\
$\alpha_R$ & Non-interfering & $|\bm{P}|>0$\\
$F_{RR'}$ & Interference $R-R'$ & \\
$\alpha_{RR'}$ & Interference $R-R'$ & $|\bm{P}|>0$\\
$\mathcal{H}^R$ & Interference $R-R'-R''$ & $|\bm{P}|>0$\\
& Interference $R-S$ & $|\bm{P}|>0$\\
\midrule
$\hat{\bm{P}}$ & Non-interfering & Parity-violation\\
$|\bm{P}|$ & Interference $R-R'-R''$ & Parity-violation\\
& Interference $R-S$ & Parity-violation\\
\bottomrule
\end{tabular}
\caption{Summary of the \Lcpkpi decay rate study. It is reported which part of the decay rate is necessary to get information on the model parameters under which conditions. See text for details. \label{tab:summary_decay_rate}}
\end{table}

\section{Toy Amplitude Fit}
\label{sec:toy_fit}
A toy amplitude maximum-likelihood fit is built to cross-check the results of the analytical study of the \Lcpkpi decay rate presented in Section~\ref{sec:decay_rate_study}. We choose a toy amplitude model which satisfy the conditions summarized in Section~\ref{sec:decay_rate_summary}: it consists of three resonances, one per decay channel, with resonance parameters and helicity couplings chosen in order to produce significant parity-violation and interference effects. Resonance lineshapes are taken to be relativistic Breit-Wigner functions~\cite{BreitWigner},
\begin{equation}
\mathcal{R}_{\rm BW}(m^2) = \frac{1}{m^2_0 - m^2 -im_0\Gamma_0}
\end{equation}
characterised by mass $m_0$ and width $\Gamma$ parameters. The following spin-parity $J^P$ assignments are considered: $K^*(1^+)$, $\Lz^*(1/2^-)$ and $\Deltares^*(1/2^-)$.

The amplitude fit code is based on a version of the TensorFlowAnalysis package~\cite{TFA} adapted to five-dimensional phase space three-body amplitude fits~\cite{Marangotto:2713231}; this package depends on the machine-learning framework TensorFlow~\cite{tensorflow2015-whitepaper} interfaced with MINUIT minimisation~\cite{James:1975dr} via the ROOT package~\cite{Brun:1997pa}.

A set of 500'000 Monte Carlo pseudo-data has been generated according to the toy amplitude model, with the set of parameters reported in Table~\ref{tab:toy_fit_pol}. For computational reasons the helicity couplings are defined relatively to a reference one, $\mathcal{H}^{K^*}_{1/2,0}$, whose value is fixed to 1. The normalisation of the decay rate is ensured computing its integral as a function of the fit parameters. The pseudo-data sample is then fit with the same toy amplitude model leaving helicity couplings (but the reference one), polarisation components and resonance masses and widths as free fit parameters. The starting and final values of the fit parameters are reported in Table~\ref{tab:toy_fit_pol}, in which uncertainties are computed from the Hessian matrix associated to the maximum-likelihood fit. The starting point in parameter space is chosen to be far enough from the generated one, so that the possibility to obtain the generated values just by chance is negligible.

\begin{table}
\centering
\begin{tabular}{lcccc}
\toprule
Parameter & Generated value & Fitted value & Uncertainty & Starting value\\
\midrule
$K^*$\\
$\mathcal{H}_{1/2,0}$ & $1$\\
$\mathcal{H}_{1/2,-1}$ & $0.5 + 0.5i$ & $0.482 + 0.4956i$  & $0.012$; $0.0087$ & 1\\
$\mathcal{H}_{-1/2,1}$ & $i$ & $-0.019 + 1.0047i$ & 0.019; 0.0088 & 1 \\
$\mathcal{H}_{-1/2,0}$ & $- 0.5 - 0.5i$ & $-0.480 - 0.526i$ & 0.014; 0.011 & 1 \\
$m(\gev)$ & 0.9 & 0.89980 & 0.00042 & 1.1\\
$\Gamma(\gev)$ & 0.2 & 0.1984 & 0.0014 & 0.1\\
\midrule
$\Lz^*$\\
$\mathcal{H}_{-1/2,0}$ & $i$ & $-0.036 + 1.009i$ & 0.017; 0.014 & 1\\
$\mathcal{H}_{1/2,0}$ & $0.8 - 0.4i$ & $0.811 - 0.375i$ & 0.011; 0.013 & 1\\
$m(\gev)$ & 1.6 & 1.60129 & 0.00069 & 1.8\\
$\Gamma(\gev)$ & 0.2 & 0.2014 & 0.0015 & 0.3\\
\midrule
$\Deltares^*$\\
$\mathcal{H}_{-1/2,0}$ & $0.6 - 0.4i$ & $0.625 -0.398i$ & 0.011; 0.011 & 1 \\
$\mathcal{H}_{1/2,0}$ & $0.1i$ & $0.0034 + 0.1191i$ & 0.0066; 0.0070 & 1 \\
$m(\gev)$ & 1.4 & 1.3994 & 0.0012 & 1.6\\
$\Gamma(\gev)$ & 0.2 & 0.2064 & 0.0023 & 0.1\\
\midrule
$\bm{P}$\\
$P_z$ & 0.5 & 0.5029 & 0.0038 & 0\\
$P_x$ & 0. & -0.0029 & 0.0036 & 0\\
$P_y$ & 0. & 0.0015 & 0.0036 & 0\\
\bottomrule
\end{tabular}
\caption{Toy amplitude fit study for non-zero polarisation ($P_z = 0.5$). For helicity couplings, the two uncertainties are associated to their real and imaginary part separately. \label{tab:toy_fit_pol}}
\end{table}

\begin{table}
\centering
\begin{tabular}{lcccc}
\toprule
Parameter & Generated value & Fitted value & Uncertainty & Starting value\\
\midrule
$K^*$\\
$\mathcal{H}_{1/2,0}$ & $1$\\
$\mathcal{H}_{1/2,-1}$ & $0.5 + 0.5i$ & $1.72 - 0.22i$  & $0.12$; $0.17$ & 1\\
$\mathcal{H}_{-1/2,1}$ & $i$ & $0.732 + 0.27i$ & 0.091; 0.29 & 1 \\
$\mathcal{H}_{-1/2,0}$ & $- 0.5 - 0.5i$ & $-1.01 - 1.29i$ & 0.16; 0.13 & 1 \\
$m(\gev)$ & 0.9 & 0.90060 & 0.00061 & 1.1\\
$\Gamma(\gev)$ & 0.2 & 0.2002 & 0.0019 & 0.1\\
\midrule
$\Lz^*$\\
$\mathcal{H}_{-1/2,0}$ & $i$ & $-0.18 + 0.726i$ & 0.13; 0.053 & 1\\
$\mathcal{H}_{1/2,0}$ & $0.8 - 0.4i$ & $1.65 - 1.09i$ & 0.12; 0.17 & 1\\
$m(\gev)$ & 1.6 & 1.6016 & 0.0011 & 1.8\\
$\Gamma(\gev)$ & 0.2 & 0.2006 & 0.0017 & 0.3\\
\midrule
$\Deltares^*$\\
$\mathcal{H}_{-1/2,0}$ & $0.6 - 0.4i$ & $0.667 -0.809i$ & 0.083; 0.075 & 1 \\
$\mathcal{H}_{1/2,0}$ & $0.1i$ & $-0.419 - 0.240i$ & 0.067; 0.059 & 1 \\
$m(\gev)$ & 1.4 & 1.3969 & 0.0019 & 1.6\\
$\Gamma(\gev)$ & 0.2 & 0.2037 & 0.0040 & 0.1\\
\midrule
$\bm{P}$\\
$P_z$ & 0 & -0.0038 & 0.0034 & 0\\
$P_x$ & 0 & -0.0005 & 0.0035 & 0\\
$P_y$ & 0 & 0.0005 & 0.0034 & 0\\
\bottomrule
\end{tabular}
\caption{Toy amplitude fit study for zero polarisation. For helicity couplings, the two uncertainties are associated to their real and imaginary part separately.\label{tab:toy_fit_zero_pol}}
\end{table}

\begin{figure}[h]
\centering
\includegraphics[width=\textwidth]{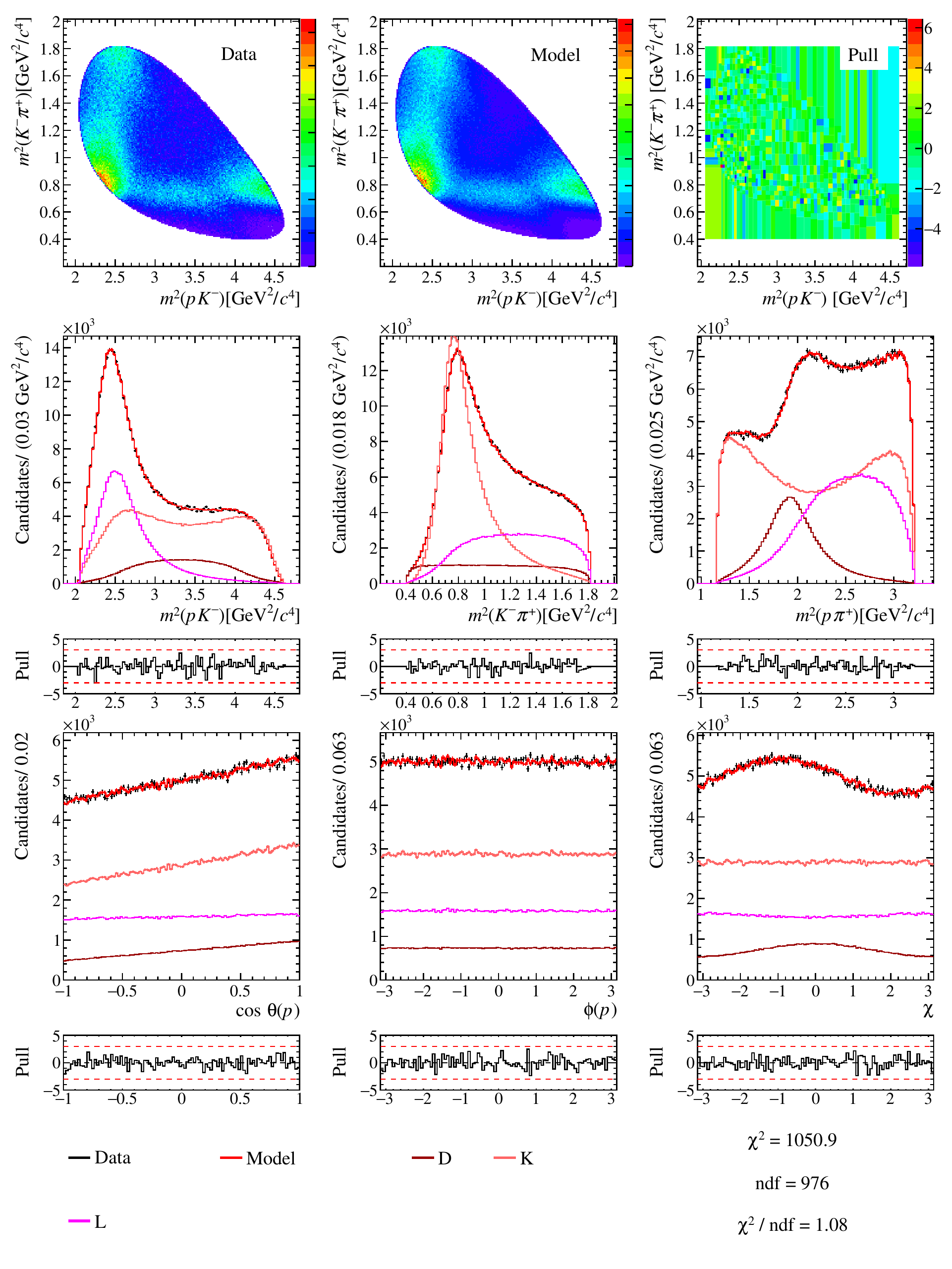}
\caption{Comparison between Monte Carlo pseudo-data (``Data'') and fitted toy amplitude model (``Model'') projections for non-zero polarisation ($P_z = 0.5$), summarized in Table~\ref{tab:toy_fit_pol}.\label{fig:Plot_P_500k}}
\end{figure}

The fit results demonstrate that the amplitude fit is able to measure simultaneously all the amplitude model parameters: complex helicity couplings, polarisation degree and direction and resonance masses and widths, as predicted by the analytical study of the \Lcpkpi decay rate. Indeed, all the parameters are found to be compatible within twice the range set by the computed statistical uncertainties. The comparison between pseudo-data and amplitude model projections is displayed in Figure~\ref{fig:Plot_P_500k}.

\begin{figure}[h]
\centering
\includegraphics[width=\textwidth]{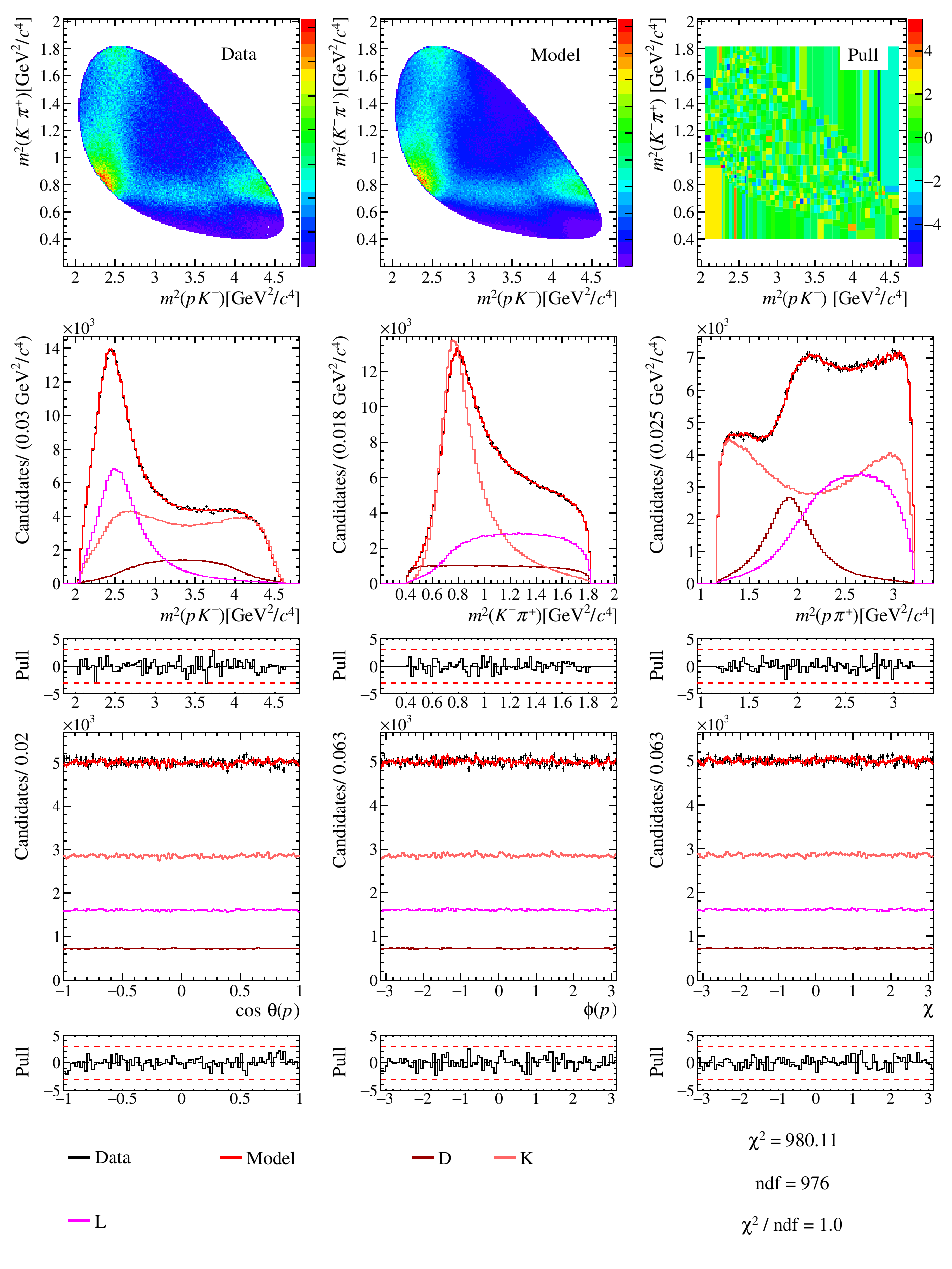}
\caption{Comparison between Monte Carlo pseudo-data (``Data'') and fitted toy amplitude model (``Model'') projections for zero polarisation, summarized in Table~\ref{tab:toy_fit_pol}.\label{fig:Plot_zeroP_500k}}
\end{figure}

To highlight the importance of having a non-zero polarisation for extracting full information from \Lcpkpi decay distributions we repeat the toy amplitude fit, exactly in the same way, but for pseudo-data generated for zero polarisation. The results are reported in Table~\ref{tab:toy_fit_zero_pol}: while the null polarisation and resonance parameters are correctly retrieved, the helicity coupling values are far from the generated ones; their associated uncertainties, much larger than those characterising the previous fit, suggest that they are not fully constrained by the pseudo-data distributions. The comparison between pseudo-data and amplitude model projections is displayed in Figure~\ref{fig:Plot_zeroP_500k}.

What about the extraction of the fit fractions, which, according to Eq.~\eqref{eq:Lcpkpi_decay_rate_non_int} should be measurable even in absence of polarisation? Due to the way the helicity couplings are defined in the amplitude fit, we should not check whether the sum of the helicity coupling moduli is retrieved, but rather the relative sum with respect to the other resonances. Indeed, from the final helicity coupling values reported in Table~\ref{tab:toy_fit_zero_pol} it is easy to check that the $F_R$ values are compatible with the generated ones but for an overall factor $\approx 2.5$, which is absorbed in the amplitude model normalisation.

\section{Conclusions}
\label{sec:conclusions}
We perform an analytical study of the \Lcpkpi decay to understand which information can be extracted from its decay distributions. This study holds as well for any three-body decay with spin structure $1/2 \to 1/2 \text{ } 0 \text{ } 0$, having a first parity-violating decay to an intermediate state subsequently decaying via parity-conserving interactions.

Considering an amplitude model for the \Lcpkpi decay written in the helicity formalism we show which parameters of the amplitude model can be measured from which part of the decay rate. We demonstrate how the presence of significant interference effects together with a non-negligible \Lc polarisation allows the simultaneous measurement of all the amplitude model parameters, including complex helicity couplings, and the polarisation degree and direction. We highlight the interplay between amplitude model and polarisation degree: significant resonance interference effects are needed to measure the second; while a non-zero polarisation allows to fully determine the decay model.

The analytical study has been cross-checked numerically by means of toy amplitude fits on Monte Carlo pseudo-data.

According to the ongoing studies of \Lcpkpi decays at LHCb~\cite{Marangotto:2713231}, this decay has all the features needed for the complete determination of its amplitude model parameters. The full determination of the \Lcpkpi amplitude model opens the possibility to use a \Lcpkpi decay amplitude model for the different applications presented in Section~\ref{sec:intro}, ranging from New Physics searches to low-energy QCD studies. In particular the \Lcpkpi decay can be used as an absolute polarimeter for the \Lc baryon.

\section*{Data Availability}
The articles used to support the findings of this study are
included within the article and are cited at relevant places
within the text as references.

\section*{Conflicts of Interest}
The author declares that there are no conflicts of interest.

\section*{Funding Statement}
This work was supported by the ERC Consolidator Grant SELDOM G.A. 771642.

\section*{Acknowledgements}
The initial idea for the present study arose during a discussion with Andrzej Kupsc at the Issues in Baryon Spectroscopy workshop organised by MIAPP in October 2019; I thank him and the organisers of the workshop. I also thank my colleagues Louis Henry, Fernando Mart\'{i}nez Vidal and Nicola Neri for the effort shared towards \Lcpkpi amplitude analysis and polarisation measurements in LHCb.

\appendix

\section{Explicit $d$-Matrices Expressions}
\label{sec:d_matrix_expressions}
Here, we report the explicit $d$-matrices expressions employed throughout the article.

The Wigner $d$-matrix for spin 1/2 is
\begin{equation}
d^{1/2}_{m',m}(\theta) = \left(
\begin{array}{ccc}
\cos\frac{\theta}{2} & & -\sin\frac{\theta}{2}\\[2ex]
\sin\frac{\theta}{2} & & \cos\frac{\theta}{2}\\
\end{array}
\right).
\label{eq:d_matrix_1/2}
\end{equation}
The Wigner $d$-matrix elements $m,m'=\pm 1/2$ for semi-integer spin $s$ can be written in the form:
\begin{equation}
d^{S}_{m',m}(\theta) = f(S,\theta) d^{1/2}_{m',m}(\theta).
\label{eq:d_matrix_s_1/2}
\end{equation}
For instance,
\begin{align}
f(1/2,\theta)&=1\nonumber\\
f(3/2,\theta)&=\frac{1}{2}(3\cos\theta-1).
\end{align}

\section{Projections of the \Lcpkpi Toy Amplitude Model}
\label{sec:projections}
\begin{figure}
\centering
\includegraphics[width=\textwidth]{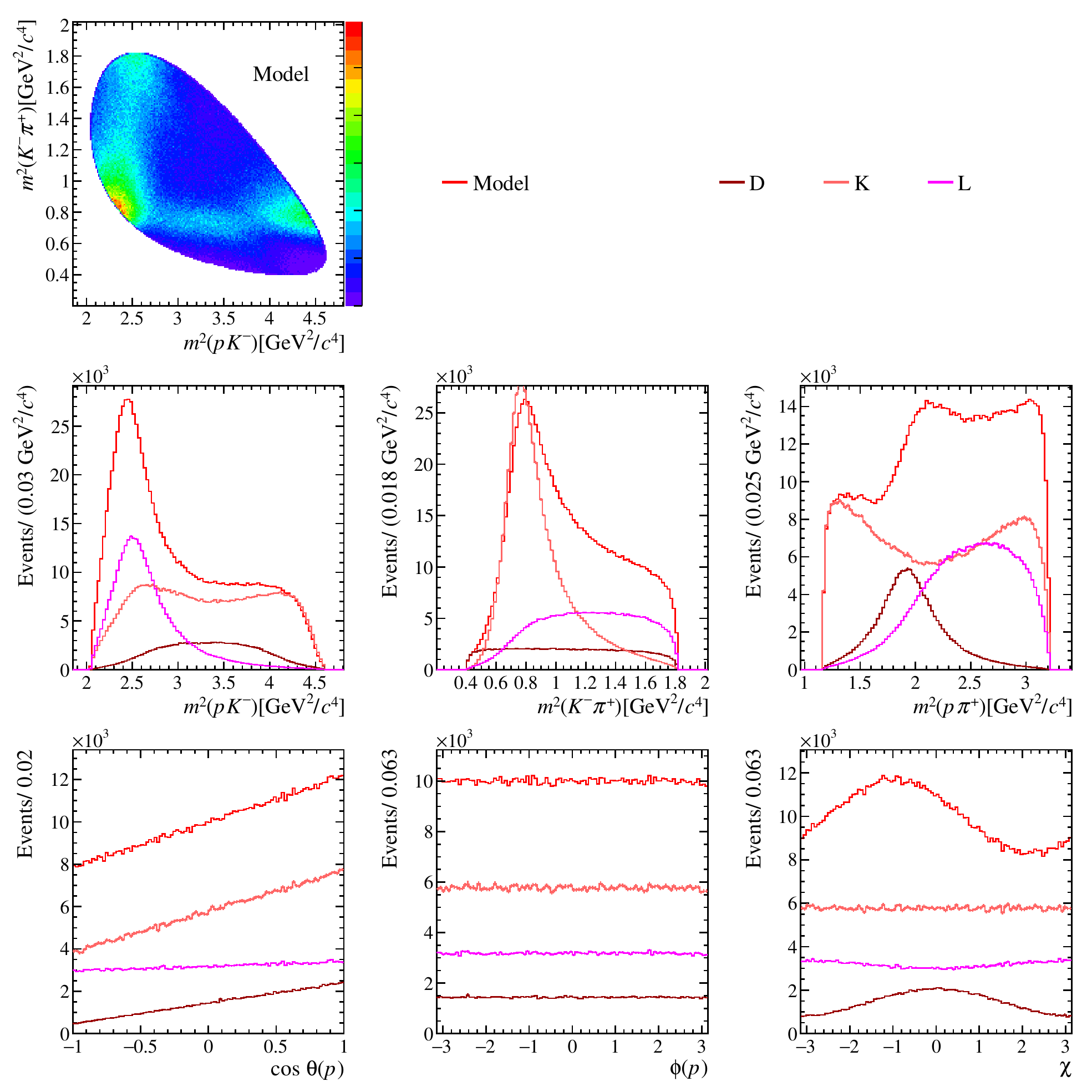}
\caption{\Lcpkpi toy amplitude model projections for $P_z=1$, generated using one million Monte Carlo events.\label{fig:Test_P_1_dp}}
\end{figure}

\begin{figure}
\centering
\includegraphics[width=\textwidth]{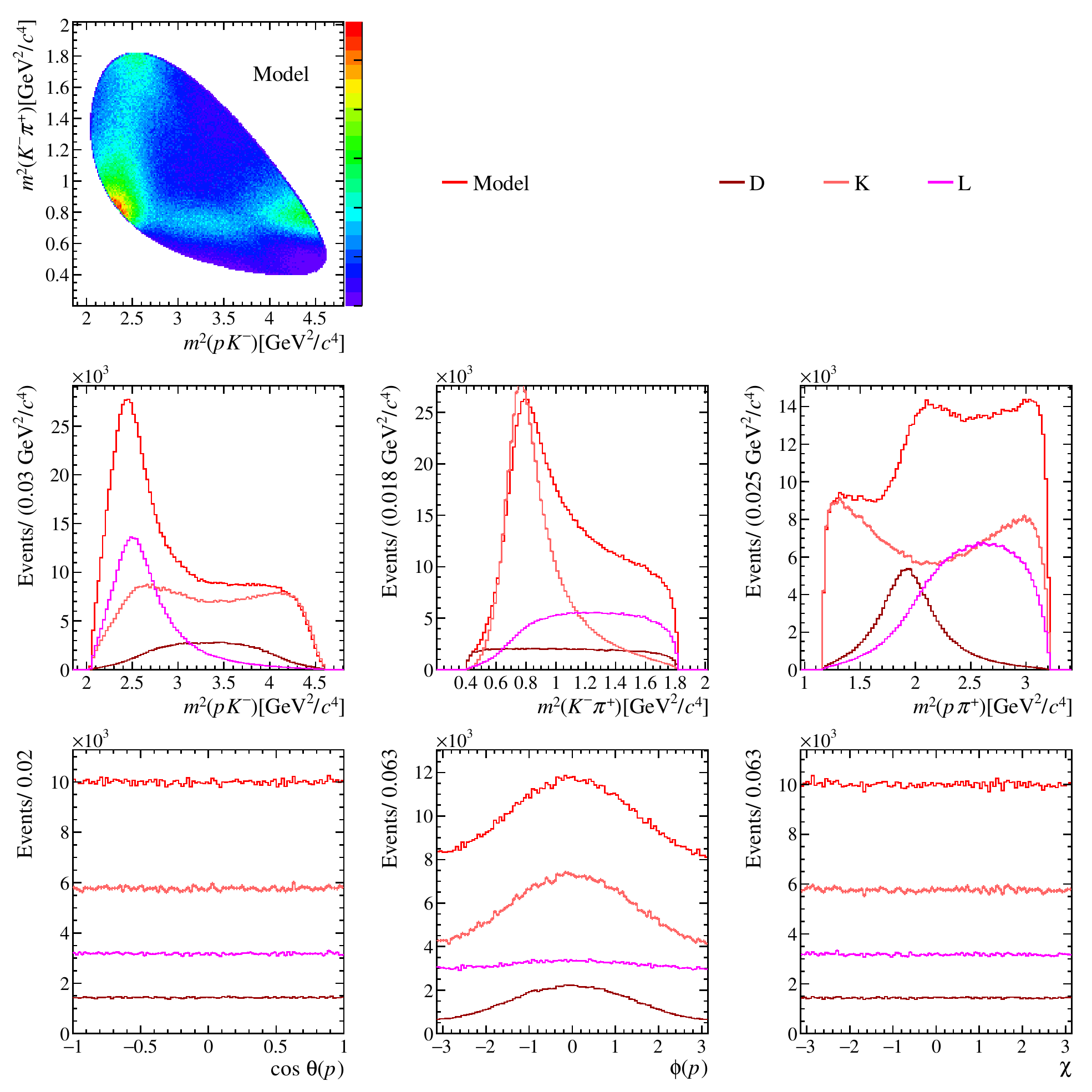}
\caption{\Lcpkpi toy amplitude model projections for $P_x=1$, generated using one million Monte Carlo events.\label{fig:Test_Px_1_dp}}
\end{figure}

\begin{figure}
\centering
\includegraphics[width=\textwidth]{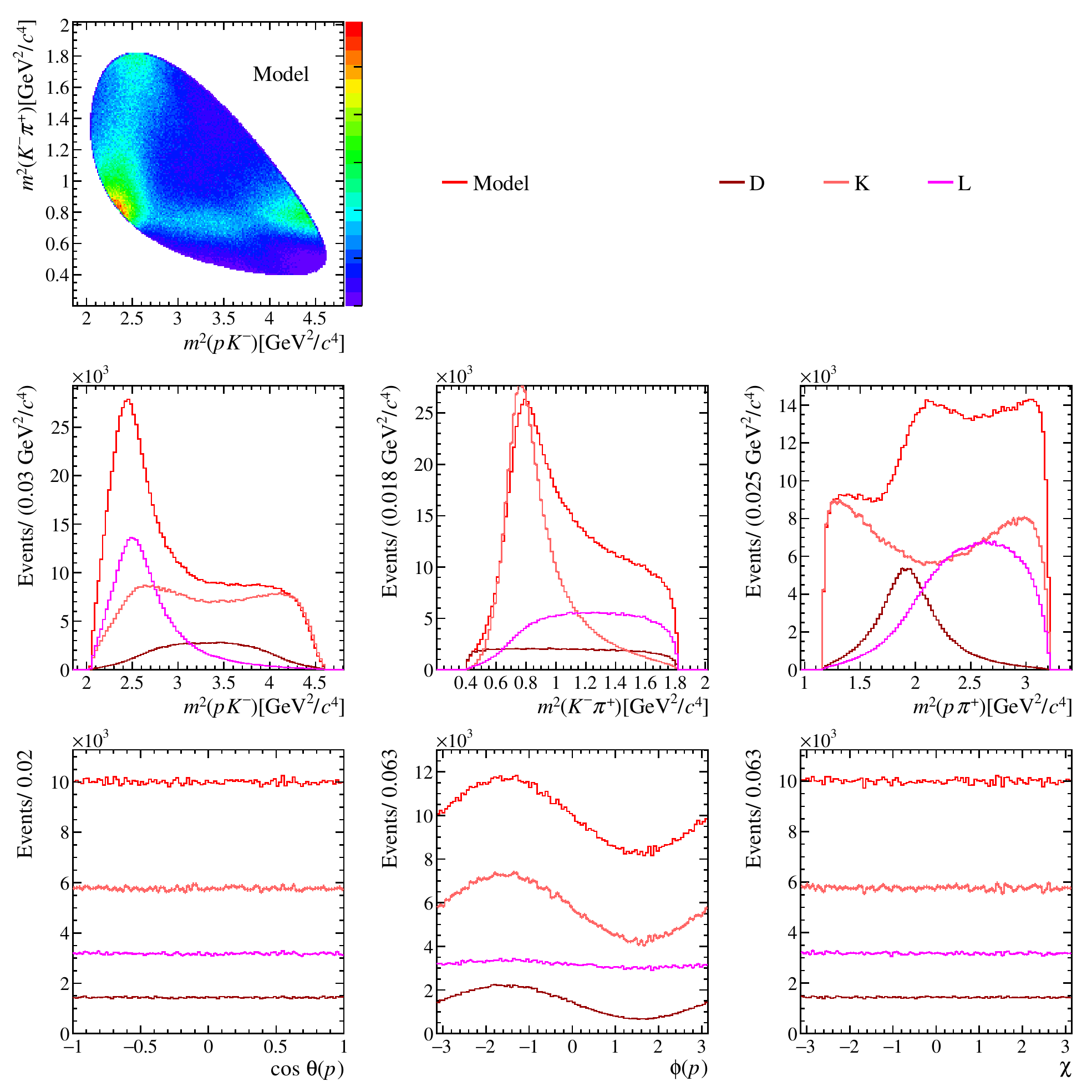}
\caption{\Lcpkpi toy amplitude model projections for $P_y=1$, generated using one million Monte Carlo events.\label{fig:Test_Py_1_dp}}
\end{figure}

We consider the projections of the \Lcpkpi toy amplitude model described in Section~\ref{sec:toy_fit} for different \Lc polarisation components: $P_z$ in Figure~\ref{fig:Test_P_1_dp}, $P_x$ in Figure~\ref{fig:Test_Px_1_dp} and $P_y$ in Figure~\ref{fig:Test_Py_1_dp}.

The orientation angle projections follow the dependencies predicted in Section~\ref{sec:general_study_pol}: a $P_z$ component produces a linear $\cos\theta_p$ and a sinusoidal $\chi$ dependence, while a $P_x$ ($P_y$) component gives rise to a cosine (minus sine) dependence in $\phi_p$. Invariant mass distributions are independent on the polarisation for rotational invariance as explained in Section~\ref{sec:properties_polarised_decay_rate}.

\clearpage

\addcontentsline{toc}{section}{References}
\setboolean{inbibliography}{true}
\bibliographystyle{LHCb}
\bibliography{biblio}

\end{document}

%% file: thesis-symbols.tex
\def\Lb      {{\ensuremath{\Lz^0_\bquark}}\xspace}

\def\Lc      {{\ensuremath{\Lz^+_\cquark}}\xspace}

\newcommand{\tevc}{\ensuremath{{\mathrm{\,Te\kern -0.1em V\!/}c}}\xspace}
\newcommand{\tevtevcccc}{\ensuremath{{\mathrm{\,Te\kern -0.1em V^2\!/}c^4}}\xspace}
\newcommand{\gevgevcc}{\ensuremath{{\mathrm{\,Ge\kern -0.1em V^2\!/}c^2}}\xspace}
\newcommand{\tevtevcc}{\ensuremath{{\mathrm{\,Te\kern -0.1em V^2\!/}c^2}}\xspace}

\def\s0z   {{\ensuremath{ s_{0,z} }}\xspace}

\newcommand{\Lcpkpi}{\ensuremath{\Lc\to pK^-\pi^+}\xspace}

\newcommand{\mqpk}{\ensuremath{m^2_{pK^-}}\xspace}
\newcommand{\mqkpi}{\ensuremath{m^2_{K^-\pi^+}}\xspace}
\newcommand{\mqppi}{\ensuremath{m^2_{p\pi^+}}\xspace}

\newcommand{\re}{\ensuremath{\mathrm{Re}\hspace{2pt}}}
\newcommand{\im}{\ensuremath{\mathrm{Im}\hspace{2pt}}}

%% file: preamble.tex
\usepackage{xspace} 
\usepackage{caption} 

\usepackage{graphicx}  
\usepackage{color}
\usepackage{colortbl}

\usepackage{amsmath} 
\usepackage{amssymb}
\usepackage{amsfonts}
\usepackage{upgreek} 

\input{lhcb-symbols-def} 

\usepackage{cite} 
\usepackage{mciteplus}

%% file: lhcb-symbols-def.tex
\usepackage{xspace} 
\usepackage{upgreek}

\def\MagUp {\mbox{\em Mag\kern -0.05em Up}\xspace}

\ifthenelse{\boolean{uprightparticles}}%
{

 \def\PDelta      {\ensuremath{\Delta}\xspace}                 
 \def\PXi      {\ensuremath{\Xi}\xspace}                 
 \def\PLambda      {\ensuremath{\Lambda}\xspace}                 
 \def\PSigma      {\ensuremath{\Sigma}\xspace}                 
 \def\POmega      {\ensuremath{\Omega}\xspace}                 
 \def\PUpsilon      {\ensuremath{\Upsilon}\xspace}                 
 
 \def\PB      {\ensuremath{\mathrm{B}}\xspace}                 
                  
 \def\PD      {\ensuremath{\mathrm{D}}\xspace}

 \def\PK      {\ensuremath{\mathrm{K}}\xspace}

 \def\Pb      {\ensuremath{\mathrm{b}}\xspace}                 
 \def\Pc      {\ensuremath{\mathrm{c}}\xspace}

 \def\Pi      {\ensuremath{\mathrm{i}}\xspace}

}
{

 \mathchardef\PDelta="7101
 \mathchardef\PXi="7104
 \mathchardef\PLambda="7103
 \mathchardef\PSigma="7106
 \mathchardef\POmega="710A
 \mathchardef\PUpsilon="7107
                  
 \def\PB      {\ensuremath{B}\xspace}                 
                  
 \def\PD      {\ensuremath{D}\xspace}

 \def\PK      {\ensuremath{K}\xspace}

 \def\Pb      {\ensuremath{b}\xspace}                 
 \def\Pc      {\ensuremath{c}\xspace}

 \def\Pi      {\ensuremath{i}\xspace}

}

\def\cquark    {{\ensuremath{\Pc}}\xspace}

\def\bquark    {{\ensuremath{\Pb}}\xspace}

  \def\Kbar    {{\kern 0.2em\overline{\kern -0.2em \PK}{}}\xspace}

\def\KorKbar    {\kern 0.18em\optbar{\kern -0.18em K}{}\xspace}

  \def\Dbar    {{\kern 0.2em\overline{\kern -0.2em \PD}{}}\xspace}

\def\DorDbar    {\kern 0.18em\optbar{\kern -0.18em D}{}\xspace}

\def\Bbar    {{\ensuremath{\kern 0.18em\overline{\kern -0.18em \PB}{}}}\xspace}

\def\BorBbar    {\kern 0.18em\optbar{\kern -0.18em B}{}\xspace}

  \def\Y#1S{\ensuremath{\PUpsilon{(#1S)}}\xspace}

\def\Deltares    {{\ensuremath{\PDelta}}\xspace}

\def\Lz          {{\ensuremath{\PLambda}}\xspace}
\def\Lbar        {{\ensuremath{\kern 0.1em\overline{\kern -0.1em\PLambda}}}\xspace}
\def\LorLbar    {\kern 0.18em\optbar{\kern -0.18em \PLambda}{}\xspace}

\def\Lb      {{\ensuremath{\Lz^0_\bquark}}\xspace}

\def\Lc      {{\ensuremath{\Lz^+_\cquark}}\xspace}

\def\to                 {\ensuremath{\rightarrow}\xspace}

\def\C#1      {\ensuremath{\mathcal{C}_{#1}}\xspace}                       
\def\Cp#1     {\ensuremath{\mathcal{C}_{#1}^{'}}\xspace}                    
\def\Ceff#1   {\ensuremath{\mathcal{C}_{#1}^{\mathrm{(eff)}}}\xspace}        
\def\Cpeff#1  {\ensuremath{\mathcal{C}_{#1}^{'\mathrm{(eff)}}}\xspace}       
\def\Ope#1    {\ensuremath{\mathcal{O}_{#1}}\xspace}                       
\def\Opep#1   {\ensuremath{\mathcal{O}_{#1}^{'}}\xspace}                    

\newcommand{\tev}{\ensuremath{\mathrm{\,Te\kern -0.1em V}}\xspace}
\newcommand{\gev}{\ensuremath{\mathrm{\,Ge\kern -0.1em V}}\xspace}
\newcommand{\mev}{\ensuremath{\mathrm{\,Me\kern -0.1em V}}\xspace}
\newcommand{\kev}{\ensuremath{\mathrm{\,ke\kern -0.1em V}}\xspace}
\newcommand{\ev}{\ensuremath{\mathrm{\,e\kern -0.1em V}}\xspace}
\newcommand{\gevc}{\ensuremath{{\mathrm{\,Ge\kern -0.1em V\!/}c}}\xspace}
\newcommand{\mevc}{\ensuremath{{\mathrm{\,Me\kern -0.1em V\!/}c}}\xspace}
\newcommand{\gevcc}{\ensuremath{{\mathrm{\,Ge\kern -0.1em V\!/}c^2}}\xspace}
\newcommand{\gevgevcccc}{\ensuremath{{\mathrm{\,Ge\kern -0.1em V^2\!/}c^4}}\xspace}
\newcommand{\mevcc}{\ensuremath{{\mathrm{\,Me\kern -0.1em V\!/}c^2}}\xspace}

\def\gsim{{~\raise.15em\hbox{$>$}\kern-.85em
          \lower.35em\hbox{$\sim$}~}\xspace}
\def\lsim{{~\raise.15em\hbox{$<$}\kern-.85em
          \lower.35em\hbox{$\sim$}~}\xspace}

\def\tell1  {TELL1\xspace}
\def\ukl1   {UKL1\xspace}

\newcommand{\eg}{\mbox{\itshape e.g.}\xspace}
\newcommand{\ie}{\mbox{\itshape i.e.}\xspace}